\documentclass[12pt,preprint]{aastex}

\usepackage{graphicx}

\newcommand{\kms}{km~s$^{-1}$}

\newcommand{\etal}{et al.}
\newcommand{\gal}{$\alpha$}

\newcommand{\gla}{$\lambda$}
\newcommand{\fuse}{{\it FUSE}}
\newcommand{\hst}{{\it HST}}
\shorttitle{TW Hya}
\shortauthors{Dupree et al.}

\begin{document}
\slugcomment{May 5, 2014}
\title{Structure and Dynamics of the Accretion Process and Wind \\
 in TW Hya
 \thanks{Data presented herein were 
 obtained at the W. M. Keck 
 Observatory, which is operated as a scientific partnership among the 
California Institute of Technology, the University of California, and the 
 National Aeronautics and Space Administration. The Observatory was made
 possible by the generous financial support of the W. M. Keck
 Foundation. Infrared spectra were taken 
at the Gemini 
Observatory, which is operated by the 
Association of Universities for Research in Astronomy, Inc., 
under a cooperative agreement 
with the NSF on behalf of the Gemini partnership: the 
National Science Foundation (United States), formerly the Science and 
Technology Facilities Council (United Kingdom), the 
National Research Council (Canada), CONICYT (Chile), 
the Australian Research Council (Australia), 
Minist\'{e}rio da Ci\^{e}ncia e Tecnologia (Brazil) 
and Ministerio de Ciencia, 
Tecnolog\'{i}a e Innovaci\'{o}n Productiva (Argentina).
This paper also includes spectra gathered with the 6.5-meter Magellan
Telescope/CLAY  
located at Las Campanas Observatory, Chile.
Additional spectra were obtained at  the  1.5m Tillinghast Telescope
 at the Fred Lawrence Whipple Observatory of the Smithsonian
Astrophysical Observatory.}}

\author{A. K. Dupree, 
N. S. Brickhouse, S. R. Cranmer,  P. Berlind\altaffilmark{2},  Jay Strader\altaffilmark{3}}
\affil{Harvard-Smithsonian Center for Astrophysics, Cambridge, MA
  02138, USA}

\author{Graeme H. Smith}
\affil{University of California Observatories/Lick Observatory,
  University of California,\\Santa Cruz, CA 95064} 
\altaffiltext{2}{Fred L. Whipple Observatory, Amado, AZ}
\altaffiltext{3}{Michigan State University, East Lansing, MI}

\begin{abstract}
Time-domain spectroscopy of the classical accreting T Tauri star, TW Hya, 
covering a decade and spanning the far UV to the
near-infrared spectral regions can identify the radiation sources, the
atmospheric structure produced by accretion, and properties of the
stellar wind. On time scales from
days to years,  substantial changes occur in emission line
profiles and line strengths.  Our extensive time-domain spectroscopy 
suggests that the broad near-IR,  
optical, and far-uv 
emission lines, centered on the star,  originate in a turbulent post-shock
region and can undergo scattering by the overlying stellar wind as
well as some absorption from infalling material.  Stable absorption features appear
in H\gal, apparently caused by
an accreting column silhouetted  in the stellar wind. Inflow of material 
onto the star is revealed by  the near-IR \ion{He}{1} 10830\AA\ line,  and its 
free-fall velocity correlates inversely with the strength of the 
post-shock emission, consistent
with a dipole accretion model. 
However, the predictions of hydrogen line profiles based on
accretion stream models are not well-matched by these observations.
Evidence of an accelerating warm to  hot stellar wind 
is shown by the near-IR \ion{He}{1} 
line, and emission profiles of  \ion{C}{2}, \ion{C}{3}, \ion{C}{4}, \ion{N}{5},  and
\ion{O}{6}. The outflow of material changes substantially
in both   speed and opacity in the yearly sampling of the
near-IR \ion{He}{1} line over a decade. 
Terminal outflow velocities that range from 200 \kms\ to almost 400 \kms\ in \ion{He}{1}
appear to be directly related to the amount of post-shock emission,
giving evidence for an accretion-driven stellar wind.
Calculations of the emission from realistic post-shock regions are needed.

\end{abstract}

\keywords{stars: individual (TW Hydrae) - stars:pre-main sequence -
stars: variables: T Tauri, Herbig Ae/Be - stars: winds, outflows -
Accretion, accretion disks}

\section{Introduction}

TW Hya (CD $-$34 7151, TWA 1, HIP 53911) remains arguably
the closest accreting T~Tauri Star (Wichmann \etal\ 1998),
and is oriented with its rotation axis almost along our
line of sight which sets the surrounding accretion disk approximately
in the plane of the sky (Krist \etal\ 2000; Qi \etal\ 2004). These
characteristics make TW Hya a subject of intensive study
at all wavelengths because it is bright and  the accretion process  
is directly accessible by this polar orientation (Donati \etal\
2011; Johnstone \etal\ 2014).

TW Hya has been a frequent target of intensive study of optical
emission lines  along with several other  
bright T Tauri stars:  BP Tau (Gullbring \etal\ 1996);    
DF Tau (Johns-Krull \& Basri 1997); DQ Tau (Basri et al. 1997); 
DR Tau (Alencar \etal\ 2001);  and RW Aur (Alencar \etal\ 2005)  among them.  These 
studies reveal the intrinsic variability of the optical lines.    These stars
have larger inclinations than TW Hya, [20$^\circ$ for DR Tau (Schegerer \etal\ 2009);
 39$^\circ$ for BP Tau (Guilloteau et al. 2011); the star RW Aur A is inclined by
37$^\circ$; others are not known].  The uniqueness of TW Hya resides in the 
imaging of the surrounding disk in the infrared (Krist \etal\ 2000)  
and in the sub-millimeter range (Qi \etal\ 2004) revealing its face-on 
orientation.  Thus the interesting polar region where
accretion is ongoing can be viewed directly.  This paper reports spectral  
sequences not only from the optical region, but the near-infrared and ultraviolet as well.
The broad wavelength coverage spans accretion phenomena, the presence of winds,
and the intrinsic chromosphere/corona of the star.  

TW Hya  itself is of low mass.  Estimates range from
0.4 -- 0.8 M$_\odot$ (Batalha \etal\ 2002; Donati \etal\ 2011; Vacca
\& Sandell 2011).  The spectral type is believed to be close to
a K7 dwarf (Alencar \& Batalha 2002). A possible later spectral type
(M2.5V, Vacca \& Sandell 2011) is controversial (Andrews \etal\
2012; McClure \etal\ 2013). The inferred mass and radius 
for a M2.5V star lead to lower free-fall accretion
velocities and consequently shock temperatures lower than measured directly 
(Brickhouse \etal\ 2010). 
This star has become a fiducial object
in the astrophysics of accretion and  low mass star formation.   

Currently, the `standard model' of accretion suggests  (Hartmann 1998)  that material
from a surrounding disk is channeled by magnetic fields in an `accretion funnel'   
towards the central star forming an accretion shock, a small 
post-shock cooling zone, and a hot spot or ring on or near
the stellar surface.  The source of optical emission is assigned 
to the accreting funnel flows (Hartmann \etal\ 1994; Muzerolle \etal\
2001; Kurosawa \& Romanova 2013). Alencar and Batalha (2002) in a very detailed 
analysis of 42 optical spectra of TW Hya acquired mostly  nightly over
a 1.4 year period described the behavior of several emission lines:
H$\alpha$, H$\beta$, \ion{He}{1} (\gla 5876) and Na D.  They noted both outflow 
and inflow signatures in these profiles and documented a correlation between veiling
corrected equivalent widths of major lines and veiling.  This suggests that increased
veiling $-$ believed to arise from the continuum and lines (Gahm \etal\ 2008) 
produced by the accretion hot spot (or ring) $-$ is related to the increased
equivalent width of optical emission lines.  The accretion rate and its variation 
can be  inferred directly from the \ion{Ne}{9} lines in the X-ray 
spectrum (Brickhouse \etal\ 
2012).   However, recent optical spectroscopy simultaneous with X-ray measurements
demonstrated the progression of accreted material in the post-shock
cooling zone  through the stellar
atmosphere, producing optical emission,  
followed by the heating of the photosphere, and subsequent 
enhancement of the corona (Dupree \etal\ 2012). This sequence
challenges the `standard model'  which attributes optical emission to the accretion
funnels (Muzerolle \etal\ 2001, 2005; Natta \etal\ 2004). In 
addition, over this time, a stellar wind 
becomes established and increases in strength over several days.   X-ray spectroscopy of 
TW Hya and the accompanying spectroscopic diagnostics of
density and temperature revealed   a large post-shock
volume  in the corona  (Brickhouse \etal\ 2010). 
This newly-discovered material has roughly 300 times larger volume and 30 times
more mass than the accretion shock itself, and signals the presence
of an accretion-fed corona. It appears to be a large turbulent billowing
structure in the corona. This process may be similar to the recently
identified brightenings in the solar corona following the impact of fragments
from an erupting solar filament (Reale \etal\ 2013).  Accreting 
material can  supply  large nearby 
magnetic structures and also  drive a stellar wind
(Cranmer 2009). 

While broad ultraviolet, far-UV,  and optical  emission lines from TW Hya are 
well documented (Alencar \& Batalha 2002; Herczeg \etal\ 2002;
Dupree \etal\ 2005a; Ardila et al., 2013) their origin
has remained elusive.  Intensive study of the X-ray spectrum of TW Hya
with CHANDRA coupled with simultaneous optical spectra  
strongly suggests that the source of its broad optical and 
X-ray emission arises from the turbulent postshock cooling volume
in the stellar atmosphere (Brickhouse \etal\ 2010; Dupree \etal\
2012), 
and not from `accretion funnels' approaching the star. 
The broad emission can suffer some absorption from the
accretion stream and can also be 
substantially modified by a stellar wind which appears to
be driven by the accretion process. The structure of this
wind, its characteristics as well as the amount of mass loss 
remain to be determined. These processes
have substantial implications for angular momentum loss
from the star (Matt \& Pudritz 2005; Matt \etal\ 2012; Bouvier \etal\ 2014), for the
presence of 
dust in the surrounding circumstellar
material (Alexander \etal\ 2005), and for an understanding of the accretion process
itself and its consequences.

To summarize, the recent measurements suggest a more complicated accretion
process than found in the `standard model' described above.  We envision a 
magnetically channeled accretion flow from the circumstellar disk towards the star.  
This is defined as an accretion funnel.  This flow of 
plasma accelerates, free-falling towards the star and  forms a shock - 
an abrupt increase in plasma temperature and density.  The post-shock 
plasma, a turbulent medium, appears to cool radiatively.
The subsequent effects of the accretion shock are many: a turbulent cooling region that
produces the broad ultraviolet emission,  the broad optical emission lines, and the
broad near-IR helium emission; a heated
spot or ring in the stellar photosphere that creates the veiling or filling-in 
of photospheric  lines; a large heated volume in the stellar corona detected in \ion{O}{7};
and acceleration of a stellar wind detected as absorption in H\gal, \ion{He}{1} and 
ultraviolet line profiles.

In this paper, many new sequences of 
optical, near infrared, and far ultraviolet spectra are presented to
infer the characteristics of the accretion process and wind on several time
scales.  The
spectra allow  identification of  new structures in the stellar 
atmosphere, give insight into the accretion parameters,  and probe the 
wind from the star.
Spectral observations in the far-UV, optical, and near-IR regions  
are described in Section 2;   the source of 
the H\gal\ line is discussed in Section
3. Spectral indications of the accretion funnels themselves are presented
in Section 4.  The H$\delta$ transition is compared to predictions of current MHD
models in Section 5. The variable warm wind\footnote{Here we take a warm wind 
to have a temperature of $\sim$15,000K indicative of the formation region of \ion{He}{1}; a hot
wind is considered to have a temperature of  $\sim$10$^5$ K or higher. A stellar wind, if like
the solar wind, exhibits a progression in temperature from `warm' to `hot'.}  
extending to  greater heights in the 
chromosphere than H-\gal\ is shown by the
near-infrared transition of \ion{He}{1} (Section 6).  Section 7
presents evidence  inferred from
the far ultraviolet spectra bearing on the variable post-shock
conditions. Section 8 evaluates a previous claim of wind
presence and characteristics.  Discussion and Conclusions 
occur in Section 9.

\section{Observations}

A variety of spectroscopic observations (Table 1) has been acquired 
to assess the structure and dynamics of the atmosphere of TW Hya. 
Optical spectra were obtained during many observing runs
at the Magellan-CLAY 6.5m telescope at Las Campanas Observatory,
Chile. MIKE, the double-echelle spectrograph (Bernstein \etal\ 2003)
was used with a 0.75\arcsec $\times$5\arcsec\ slit that yields  a 
2-pixel resolution element of $\lambda/\Delta\lambda \sim $30,000 on the 
red side ($\lambda\lambda$4900--9300) which contains the H$\alpha$
line and $\lambda/\Delta\lambda\sim$37,000 on the blue side ($\lambda\lambda$3200--5000) used for the higher
Balmer lines.  The IDL pipeline developed by S. Burles, R. Bernstein,
and J. X. Prochaska\footnote{See  http://web.mit.edu/\~{ }burles/www/MIKE/} 
was used to extract the spectra, and IRAF software\footnote{See {\it iraf.noao.edu}}
was  employed in the  analysis.  Additional optical  spectra were taken
at the FLWO 1.5-m telescope with the Tillinghast Reflector Echelle
Spectrograph (TRES).  TRES is a 
temperature-controlled, fiber-fed instrument with a resolving power
R$\sim$44,000. The spectra were reduced with a custom IDL pipeline.\footnote{See
  {\it www.sao.arizona.edu/FLWO/60/tres.html}}     

Near-infrared spectra were obtained with PHOENIX at Gemini-S during
two classical observing runs.  The PHOENIX setup consisted of a slit 
width of 4 pixels yielding a spectral resolution of $\sim$50,000. The
order-sorting filter J9232 was selected which spans 1.077--1.089$\mu$m 
and allows access to the \ion{He}{1} line at 1.083$\mu$m.  Standard
procedures were followed by acquiring target spectra using a nodding 
mode (A--B) with a spatial separation of 5 arcsec. Because standard
comparison lamps have a sparse wavelength pattern in this
near-infrared
region, we observed a bright K giant containing many securely
identified narrow photospheric absorption lines in order to determine the
wavelength scale.  Other near-infrared spectra were taken with NIRSPEC
(McLean \etal\ 1998, 2000) on the Keck II telescope.  The echelle
cross-dispersed
mode of NIRSPEC with the NIRSPEC-1 order-sorting filter was selected 
and the slit of 0.42\arcsec $\times$12\arcsec\ gives a nominal
spectral resolution of 23,600. The long-wavelength blocking filter was not used in order
to minimize unwanted fringing.  Internal flat-field lamps, NeArKr
arcs, dark frames, and K0 giants were used as calibration
exposures.  Spectra were taken in a standard nodding mode. 
Data reduction was carried out with the REDSPEC package 
specifically written for NIRSPEC (McLean \etal\ 2003).

Far-UV spectra of TW Hya from FUSE were taken from the MAST archive and
the extraction procedure was fully described elsewhere (Dupree \etal\ 2005a).
Some HST spectra of TW Hya obtained with STIS were also taken from the MAST archive
at STScI, and  from the CoolCAT UV spectral catalogue (Ayres 2005).\footnote{See {\it
    casa.colorado.edu/\~{ }ayres/CoolCAT/}.}  Other HST spectra came
directly from the MAST archive at STScI. Details of these spectra are
also given in Table 1.

\section {Identifying the Source of H\gal }

The H\gal\ profile is useful for probing the accretion process.  
A broad  H\gal\ line has long been considered a signal
of ongoing accretion in young stars (Bertout 1989; Hartmann \etal\ 1994;
Alencar \& Batalha 2002), and the cause of the broadening has
frequently been ascribed to `turbulence' (Alencar \& Basri 2000). In this section,
we present the most extensive study of TW Hya in the optical region.   The 
H\gal\ profile   was measured on 41 nights over  a decade 
(Figure  1),
with anywhere from 2 to 300 spectra per night.  Most of the
observations were taken on sequential days which enables the identification 
of dynamic events in a stellar atmosphere.  As has been noted almost
two decades ago (Johns-Krull \& Basri 1997), snapshot spectra 
do not provide a characteristic representation of a complex accreting 
system, and time sequences are invaluable.  The  H\gal\ profiles in TW Hya 
show dramatic changes in the strength of the emission and the 
presence of absorption features at negative (outflowing)  velocities.\footnote{Water
vapor produces an absorption feature on the positive velocity
side near $+$50 \kms.} The signature of a wind
in the H\gal\ profile has been noted previously (Hartmann 1982; Calvet \& Hartmann 1992; Gullbring \etal\ 1996; 
Alencar \&\ Basri
2000; Alencar \& Batalha 2002) but only recently has  the  hourly development of
wind absorption became  evident (Dupree \etal\ 2012).  The shape of
the emission profile  indicates material motions of gas at 
chromospheric temperatures.  Comparison of profiles in 
Figure 1 and evaluation of the flux at negative and positive velocities 
demonstrates the presence of roughly symmetric H\gal\ profiles (to within 10\%) 
centered on TW Hya for 42\% of the spectra  and a
positive flux enhancement (`red enhanced') for 52\% of the spectra. 
The  full width at half maximum (FWHM) of the lines rangew from 200 to 300
\kms. This positional symmetry is consistent with formation on the
star itself. When the  profiles appear to be   
`red-enhanced', this is  caused by substantial absorption and scattering 
(on the ``blue-side'') at negative velocities by outflowing material
forming the stellar wind. A particularly instructive sequence
occurred in 2007 Feb (Dupree
\etal\ 2012).  At that time, the H\gal\ profile was roughly
symmetric on the first night of observation, then the wind
opacity systematically increased over the following 3 nights.  

These observations demonstrate  that the H\gal\ profile
from TW Hya is inherently broad and symmetrically positioned on the
stellar velocity, but 
can be substantially absorbed by the stellar wind. In addition,
changes in the long-wavelength side of the profile (cf. Figure 1) 
can result from a combination of enhanced post-shock emission and 
changing absorption by  the accretion stream.  The pole of
TW Hya is observed directly.  It is here that  accretion occurs most
of the time (Donati \etal\ 2011),  and the shock and
post-shock cooling region can be seen. 
The source of this broad  emission has been long debated. With the 
assumption that emission
arises from the accretion stream, models  tend to predict
pointed profiles from a star observed at low inclination.    
Frequently  absorption on the red side that results from  the
infalling material  emerges from the calculations (Muzerolle
\etal\ 2001).  This makes  H\gal\
appear shifted to shorter wavelengths in contrast to most of  these
observations. 
Authors have  noted that the magnetospheric accretion models
do not adequately represent spectral profiles found in several T Tauri
stars (e.g. Alencar \& Basri 2000).  
Changes in damping parameters and the  introduction of Stark 
broadening, can manage to broaden the H\gal\ profile but the observed
profiles in TW Hya have flatter peaks and are frequently wind absorbed in
contrast to the calculated profiles (Figure
2).   

During many 
nights the observed  H\gal\  profile remains centered on the stellar radial velocity
and appears symmetric without a distinct absorption feature.  
As previously suggested (Dupree \etal\ 2012), the
broad H\gal\ lines can result naturally  from turbulent conditions in a  
post-shock cooling volume.  This identification derived from the
time study of an X-Ray accretion event which was followed by
sequential changes in emission line profiles and fluxes, veiling, and
subsequent coronal X-Ray enhancement.  The  breadth of the H\gal\ line is
slightly larger ($\sim$200$-$275 \kms), but comparable to that of   \ion{Ne}{9} 
X-ray lines arising in the shock (126 to 229 \kms)  while
slightly less than the full width of the symmetrized  far-ultraviolet 
lines (\ion{C}{3} and \ion{O}{6}) observed to 
be $\sim$~325~\kms\ in TW Hya (Dupree \etal\ 2005a).  
Moreover, the amount of material in the post-shock region inferred from
the X-ray diagnostics (Brickhouse \etal\ 2010)  as well as 
near-infrared measurements (Vacca \& Sandell 2011) exceeds by a few orders
of magnitude the material expected in an accretion column (cf. Dupree
\etal\ 2012). These facts strongly suggest that  the  symmetric H\gal\ 
emission line arises predominantly from the 
turbulent post-shock cooling region. 
It has been known for a long time that models of a
stellar chromosphere alone can not produce  strong
hydrogen emission lines (Calvet \etal\ 1984).
A symmetric H\gal\ profile centered on the stellar
radial velocity is  expected  from TW Hya because  the postshock region is in
our full view.  When a wind from the star is present,
the short wavelength side of the profile is weakened, and the
resulting profile is stronger on the long wavelength
side of the line as frequently appears in TW Hya.  Absorption by 
infalling material can affect the profile as well. 

Systems with different inclinations might be expected to produce similar
turbulently broadened profiles if the post-shock region is not obscured
by a disk.  In fact, broad H\gal\ has long been known 
(Bertout 1989; White \& Basri 2003) to signal the presence of a classical accreting
T Tauri star.    Differences in the profile may arise from wind scattering which 
would be determined by  the line-of-sight component of the 
expansion velocity. But as is evident from Figure 1, where the line-of-sight
reflects the maximum outflowing velocity from TW Hya, the wind velocity
and opacity can change substantially with time.  The \ion{He}{1} profile,
discussed in Section 6 offers an interesting test of the effects of inclination.

\section{Detection of Shadowing by Accretion Funnels}

The H\gal\ profiles also reveal structures that occur in the 
stellar wind and their behavior becomes apparent in the spectral sequences 
lasting several days. Alencar and Batalha (2002) remarked on a broad
(FWHM $\sim$125 \kms) absorption feature in H\gal\ that appeared
in nightly spectra suggesting variation in wind opacity.  
Time domain spectroscopy presented here confirms the changes in
the wind, and also reveals another feature.    Profiles
in Figure 1 show an absorption
feature that occurs at $\sim$$-$50 to $-$100 
\kms.  This feature can change in strength and velocity
from night to night, but is constant in velocity over
a single night.  Particularly instructive are the nights
following 2004 Apr 27 (Figure 3) where the absorption feature appears at $-$80 \kms, vanishes
on Apr 28, re-appears on Apr 29 at $-$150 \kms, and on Apr 30 at $-$100 \kms. Similar 
behavior occurred in 2007 Feb and 2009 Jun (Figure 1) and also in 2006
(Figure 4).   These absorption features in
the wind remain at constant velocity during the night. 
The three-night sequence from 2006 April
is presented  in  gray scale representation in Figure 5.  The first night shows
a narrow ($\sim$50 \kms) absorption feature at $-$125 \kms\ at constant velocity; this
feature becomes weaker early in night 2.  However, during the second
night of observations,  the absorption at $-$50 \kms\ at the
start of the night is replaced by an absorption feature 
at $-200$ \kms\ by the end of the night. The right panel of Figure 5
which contains a line plot of the 8 hours of observation during night 2 
illustrates this jump quite clearly. In addition, the variable infall of material
at high velocities can be seen in the profile.  The short rotation period
of TW Hya (3.57 days; Hu\'elamo \etal\ 2008) causes 
different viewing angles to occur during the 3 successive nights 
of our observations. 
The  narrow absorptions appear constant in velocity suggesting 
they are located at high latitudes on the star.  If the
features moved outward within the wind, the speeds of
$\sim$150 \kms\ would allow material to cover
many ($\sim$6) stellar radii over an 8-hour span of observations 
and this motion would also be
visible in the line profiles.  But such a shift  is not observed. 

Such  static  behavior differs  from
high-lying   stellar prominences  found in rapidly rotating
stars such as AB Dor (Collier Cameron \& Robinson 1989) and
in  rapidly rotating  T Tauri stars  (Oliveira \etal\ 2000; Skelly \etal\ 2008;
G\"unther \etal\ 2013)
where the absorption feature moves rapidly in velocity from
negative to positive (or vice versa) as the feature traverses 
the stellar disk.   The absorption features in TW Hya are stable in velocity for hours, 
and are not at all similar to the moving Discrete Absorption Components (DACs)
observed in the winds of O stars (cf. Howarth \etal\ 1995) which
generally are very broad ($\sim$400 \kms) and move at thousands
of \kms\ through the atmosphere (cf. Kaper \etal\ 1999); only    
absorption features in one Wolf-Rayet star (HD 50896)  are broad, but static (Massa
\etal\ 1995). 

{\it The static nature of these relatively narrow absorptions in TW Hya suggests
they arise in a stable cool structure located in the wind itself that
is silhouetted in the hydrogen line profile.} These could be the signature of
the accreting columns crossing the path of the wind.  In
an accelerating wind, features located at $-$50 to $-$150 \kms, would 
exist  at about 1.2-1.3 R$_\star$  (Dupree \etal\ 2008), judging 
from the chromospheric helium line
asymmetries, and above the shocked material produced
by accretion (Sacco \etal\ 2010).
The strength of the absorptions, from 0.4 to 0.6 of the local
continuum  provided by the H\gal\ emission,  indicates an 
optical depth, $\tau$ of $\sim$0.7.  Here we take
$\tau = \kappa \rho H$ where $\kappa (T, n_2/n_1)$ is the line absorption
coefficient (incorporating a dependence on temperature and the 
level population ratio of hydrogen), $\rho$ is the hydrogen density,
and $H$, the thickness of the line-forming region.  The temperature
and density structure of the `funnel flows' is uncertain. However,
the preshock electron density was determined (Brickhouse \etal\ 2010) to
be 5.8$\times$10$^{11}$ cm$^{-3}$ which implies a hydrogen density
of 5$\times$10$^{11}$ cm$^{-3}$ for a plasma with 10\% helium. For
a temperature T $\sim$7000K,  the observed 
broadening and our non-LTE calculations\footnote{The PANDORA code (Avrett \& Loeser 2008) 
was used in a spherical expanding semi-empirical  model constructed to match 
wind scattering profiles of H\gal, \ion{He}{1}, and the chromospheric density inferred from the
ultraviolet \ion{C}{2} lines.}
for chromospheric hydrogen suggest that the n$_2$/n$_1$
level populations of the H\gal\ line range from  6.4$\times$10$^{-7}$ to 1.8$\times$10$^{-8}$ 
in this temperature and density regime (Dupree \etal\ 2008).  
The scale length to produce the narrow absorption is
small, less than 0.01 R$_\star$.  Another model of the
accreting  stream (Hartmann \etal\ 1994) gives T = 7000K and the level
2 population of hydrogen as 8 $\times$10$^5$ cm$^{-3}$ near the
star, such that the inferred thickness is much smaller, $\sim$100 km.  In
either case, formation in  a narrow accretion funnel appears plausible
to account for these absorption features. 

Chromospheres, winds, and wind structures produce asymmetric lines and
absorption features that may be broad (such as a central reversal) or
narrow as discussed here. The uniqueness of these narrow `notches' 
relates to their constant velocity in the H\gal\ line profile.  A blocking
structure located in the wind could be seen at all angles where it  
 is observed against a region producing emission.   Many studies of
emission from T Tauri stars have been published but very few have the
time sampling of multiple sequential spectra that are presented here.

\section{The H$\delta$ Line}

Sophisticated models of magnetospheric accretion processes have also  
been constructed and used to predict hydrogen line profiles.
A  sequence of Balmer lines calculated by Kurosawa \&
Romanova (2013)
suggested that the H$\delta$ transition at 4101\AA\  is an optimum line to reveal persistent
red-shifted  sub-continuum absorption resulting from the inflowing accretion stream in
the `standard' model.  Absorption is expected when the
hotspot is in the line of sight of the observer as is the accretion
stream.  TW Hya is a good test of these assumptions.  Because of 
its pole-on orientation, the accretion stream and hot spot are in
the line of sight.  
Our MIKE spectra capture the pivotal H$\delta$  line, as well
as the complete Balmer series. Figure 6 shows four lines of the Balmer series
(H$\alpha$, H$\beta$, H$\gamma$, H$\delta$) taken over 4 sequential
nights in 2007 (Dupree \etal\ 2012).  During this period the H$\alpha$
line displays a slight  inflow asymmetry  (blue side stronger than red
side) early on the first night, followed by
the subsequent onset of a wind.  H$\beta$ and H$\gamma$ exhibit an absorption
feature at $\sim +$50 \kms\ which signals enhanced inflow following
the X-ray accretion event (Dupree \etal\ 2012).
However the predicted sub-continuum absorption signalling infalling material is not evident
in the H$\delta$ line.  
The period of radial velocity variation, namely 3.57 days (Hu\'elamo et
al. 2008) could cause a change in the absorption strength if
there were substantial modulation in the orientation of the hot spot
and the accretion column.  The lack of an absorption feature  in the H$\delta$ profile 
suggests that the
H$\delta$ emission  does not arise from infalling material, but must 
originate from another region.   Another sequence of H$\delta$ profiles
over 4 consecutive nights is shown in Figure 7.
The first night of the sequence shows enhanced emission with
shallow broad absorption between $+$100 to $+$200 \kms, inferred from
the line asymmetry.  This pattern may be the signal of an accretion
event.  However,
subsequent nights (July 14$-$16) reveal a symmetric profile,
indicating the
infall has lessened, and the symmetry and FHWM ($\sim$150 \kms) are   consistent
with formation in a turbulent post-shock cooling volume.  Variability
of the accretion rate is also documented by the X-ray diagnostics
(Brickhouse \etal\ 2012) which have shown a five-fold reduction in the 
rate from exposures separated by 2.7 days.   Other studies have 
measured H$\delta$ in T Tauri stars (Edwards \etal\ 1994; Petrov \etal\ 1996; 
Alencar \& Basri 2000) and the profiles in the majority of stars do not resemble 
the models presented in Kurosawa \& Romanova (2013).

\section{Variable Warm Wind in Near IR Helium} 

The broad near-IR line of \ion{He}{1} at 10830\AA,  because of
its width (FWHM $\sim$200 \kms), also appears formed in
the post-shock material.  This feature  has become 
a useful probe of the wind and infall environment of 
accreting T Tauri stars (Dupree  2003; Edwards \etal\ 2003; 
Dupree \etal\ 2005a; Edwards \etal\ 2006).  For TW Hya,
we have acquired near-IR spectra spanning 8 years  including sequences of 
consecutive nights.     A comparison of the
spectra obtained during this 8 year span is shown in Figure 8. 
Over this long time scale, practically every characteristic of 
the P Cygni profiles changes: the emission level, the wind speed and opacity,
and the appearance (or not)  of inflowing material. 
The emission strength varies by more than a factor of 2; the terminal
velocity of the wind varies from $-$200 to $-$315 \kms; the wind opacity
changes over the absorption profile; the infall terminal  velocity
varies from +270 to +363 \kms\ as does the opacity of the infalling
material. Speaking precisely,  an {\it outflow} of material is observed
in the near-infrared helium profile of TW Hya.  The escape 
velocity from the photosphere is 530 \kms\ for
TW Hya assuming M=0.8M$_\odot$ and R=1.1R$_\odot$ (Donati \etal\ 2011).  At a distance of
1 stellar radius above the surface, this value decreases to 400 \kms,
which is comparable to the observed typical terminal velocities.
It is highly likely then that the outflows observed in the near-IR
helium line and the sequential acceleration with temperature
displayed
by the UV lines in Dupree \etal\ (2005a) form a true stellar
wind.\footnote{Other studies (Batalha \etal\ 2002) suggest
  M=0.7M$_\odot$ and R=0.8R$_\odot$ 
which would indicate a comparable escape velocity (410 \kms) at 1
stellar radius above the photosphere.}

The time-domain spectroscopy revealing these changes suggests that
all T Tauri objects undoubtedly exhibit similar variability 
that snapshot studies (Edwards \etal\ 2006)  can not discern.
The extent of the variation also blurs the search for correlations of wind parameters
with geometric inclination  of the star or disk (Appenzeller \& Bertout 2013).

The helium absorption profile is broad, signalling its origin in a stellar wind, through 
its classic P Cygni shape. There
is no sign of a narrow low-velocity ($\sim$10 to 100 \kms) feature such
as identified in several spectra of other T Tauri stars (Edwards \etal\ 2006) and
conjectured to originate from the circumstellar disk. In the case of TW Hya, H$_2$
and [O I] emission arise from the disk (Herczeg \etal\ 2002, Pascucci \etal\ 2011)
since their velocities coincide with that of the star itself.  The infrared [\ion{Ne}{2}]
transition exhibits a low velocity ($-$5 \kms) outflow which suggests
the presence of a cool (1000 K) photoevaporative molecular wind (Rigliaco \etal\ 2013).  
It is hard to see how helium which is more highly excited would be produced in such a wind.  

Several sequences  of the \ion{He}{1} line over consecutive nights  are shown in 
Figure 9.  At these times minimal variation of the line profiles
occurred on a short time scale:  a small change in the emission in
2009; changes in the inflow opacity in 2010. During 2009 
and 2010 the wind absorption had constant high optical
depth (a flat-bottomed  profile) over a 50 to 100 \kms\ span, and subtle changes in
opacity would not be observable.  In contrast, the H\gal\  profile
over sequential nights reveals changes in wind opacity and absorption
troughs (cf. Fig. 1 A-E, I, K, L). In 2009, the absorption feature
at $-$50 \kms\ is also found in the H\gal\ profile from
those nights (see Figure 1). At that time, the H\gal\ line revealed
wind scattering that extends to $-$100 \kms, and the helium
wind extended to $-$200 \kms, an  indication of accelerated
expansion.  

The emission strength  exhibits substantial change.  Of course, the amount
of emission depends on the intrinsic line strength which is then modified
by the wind scattering and some absorption from infalling
material. Close inspection of the profiles suggests that
the emission itself is indeed changing.  On one occasion,  2002 May 20, the profile 
displays emission above the continuum level extending to both $-$300 and $+$300
\kms\ indicating a strong and broad intrinsic emission component. 
  
The inflow signature in helium is a particularly useful diagnostic
of the accretion process.  In a standard magnetospheric
accretion model (Brickhouse \etal\ 2012), the terminal velocity of the 
infalling material, determined by the free-fall distance from the
circumstellar disk, controls the temperature of the shock.
X-ray spectra indicate that the shock temperature, measured directly
from the ratio of forbidden plus intercombination to resonance line
fluxes of \ion{Ne}{9},  varied by a factor of 1.6 over a span of 13
days between CHANDRA pointings (Brickhouse \etal\ 2012). The shock  
temperatures vary between 1.9MK and 3.1MK. In this standard
model, the post-shock temperature is given by  $T_{post} \sim 3mv_{ff}^2/16k$, where
$v_{ff}$ is the free-fall velocity, $k$ is the Boltzmann constant 
and $m$ is the mean atomic mass
(Brickhouse \etal\ 2010).  The measured shock temperatures from 3 
CHANDRA pointings in 2007 suggest  that terminal free-fall velocities from
380 to 485 \kms\ occurred. 

In the model (Brickhouse \etal\ 2012),  when the  
free-fall velocity is low, the disk supplying the  accreting 
material originates closer in to the star. The filling factor
on the stellar surface  increases   
which implies an increase in the mass accretion
rate, and also in  the material in the post-shock
cooling zone.  A range of parameters for this model is
shown in Figure 10 ({\it left panel}).  This scenario is confirmed by the
behavior of the helium line.  {\it The amount of emission in the 
\ion{He}{1}  line appears inversely linked to the terminal velocity of 
the inflowing material} (Figure 10, {\it right panel}). Lower values of the
inflow terminal velocity lead to much stronger emission
arising in the post-shock region, which is consistent with  the higher
mass accretion rate as predicted in the dipole model.  The helium
absorption indicates lower velocities than the terminal velocities of the
models suggesting that the absorption arises from a volume above
the accretion shock.   It is also quite likely that the neutral helium is ionized 
by the X-ray emission from the shock as the helium approaches in the pre-shock
stream (Lamzin 1999; Gregory \etal\ 2007; Brickhouse \etal\ 2010), and as a result
does not reach a terminal free-fall velocity.

It is interesting
that the helium line gives evidence of  subcontinuum  absorption 
due to  inflow, when the hydrogen
lines generally do not show  such  strong  signatures of inflowing material.
On 2007 March 1, we obtained (almost) simultaneous spectra of helium
(Figure 8) and the hydrogen lines (shown as night 4 in Figure 6).  The
reversal
of the negative velocity side of the Balmer lines on to the positive 
velocity side (Figure 6) indicates weak broad absorption near $+$200
\kms\ on the H$\beta$, H$\gamma$, and H$\delta$ transitions.  However
the helium line displays substantial absorption extending to
 $+$325 \kms.  The temperature structure along the accretion 
column is not known but it appears likely that helium forms at 
higher temperatures where the hydrogen is
ionized, and these temperatures are associated with higher infalling 
velocities. This could create a higher opacity in helium with 
the ability to probe a different, potentially more extended 
accreting region than represented by  the hydrogen lines.

Because stronger emission signals enhanced accretion of material,
this material might cause an increase in both the wind opacity and
wind speed if
the accretion contributes to wind acceleration. There is a
hint of this in nine out of the ten near-IR spectra of
\ion{He}{1} (Figure 11). The
one outlier with the fastest outflow speed belongs to  the 2007 March 1 
observation discussed above.  Not only does this profile exhibit the highest
outflow velocity, but also the largest subcontinuum `blue absorption' of all
ten nights of observations. It is difficult to quantify this further
because the intrinsic profile of the helium line is unknown.  The profile 
exhibits absorption on the `red' side caused by accretion, and substantial
scattering on the `blue' side in the stellar wind.  Figure 11 shows that
increased emission in \ion{He}{1} 10830\AA\ may be linked directly to a faster
outflowing terminal velocity, giving evidence for an accretion-driven
stellar wind.

\section{Variable Post-Shock Conditions from UV lines}

The UV and far-UV lines  also exhibit  changes in 
flux as evidenced by HST and FUSE spectra. 
The expanding atmosphere of TW Hya is revealed by the asymmetric
profiles  of the major resonance lines (Dupree \etal\ 2005a), but 
substantial changes in the profiles 
are evident at different times. Figure 12 ({\it left panel}) contains
 resonance line profiles
of \ion{C}{4} (1548\AA)  and \ion{N}{5} (1239\AA) measured with
HST/STIS on 8 May 2000.  The line profiles are strikingly similar to one
another, confirming formation in the same region of the
atmosphere. Ardila \etal\ (2013) also remarked recently on this
\ion{C}{4}/\ion{N}{5} 
similarity from a much larger sample of accreting stars.  
The temperatures of formation of \ion{C}{4} and \ion{N}{5} in a collisionally ionized
plasma are similar at 1--2$\times$10$^5$K.  The profiles 
exhibit both extended emission to
positive velocities, and the characteristic sharp cutoff caused by
wind absorption at negative velocities.  More 
wind opacity is evident in the \ion{C}{4} profile than in \ion{N}{5}.  This 
is to be expected since the strength of absorption is proportional
to the value of $gf\lambda \times N_i/N_H$
where  $N_i$ is taken to be $N_C$, $N_O$, or $N_N$ giving the
appropriate elemental abundance. 
The value of $gf\lambda \times N_C/N_H$ for carbon  is larger than the similar
quantity for the \ion{N}{5} transition.  Note the extended 
wing at positive velocities on this date giving a half-maximum value
of $+$200 \kms.  Contrast these profiles 
with the  resonance lines of \ion{O}{6} and \ion{C}{3}  on 20 Feb
2003 (Figure 12, {\it right panel}) 
measured with \fuse.  Both of these
lines are also similar to one another, and their
formation temperatures are typically 8$\times$10$^4$--3$\times$10$^5$K in
equilibrium conditions. The \ion{C}{3} profile exhibits more wind
opacity than \ion{O}{6} which again can be understood from the difference
in the quantity $gf\lambda \times N_C/N_H$ which is larger for \ion{C}{3} than
for \ion{O}{6} by a factor of 2.8.  The relative amount of absorption in the wind is
in harmony with atomic physics. Thus, {\it based on these spectra, we 
conclude that the UV and far-UV lines are produced in the same
plasma region, but 
conditions in this region change dramatically from time to time.}

Intercomparison of two of the ultraviolet profiles  (Figure 13) reveals
the meaningful differences.  Excess emission on the positive velocity
side of the profile in 2000 does not occur in the 2003 observation.
The half-maximum value of the \ion{O}{6}  profile in 2003 extends only
to $+$125 \kms,
whereas \ion{C}{4} extended to $+$200 \kms\ in 2000. 
Additionally there is greater wind scattering on the negative velocity side of the
profile in  \ion{C}{4}  as compared to  \ion{O}{6}.
While  changes in the wind opacity 
between the two observations are unknown, the atomic physics of these lines
must be a first consideration.  \ion{C}{4} possesses a 17\% 
larger atomic parameter ($gf\lambda \times N_i/N_H$) than \ion{O}{6}
(where $N_i$=$N_C$ or $N_O$) which
contributes to its greater opacity.  We also do not know the
distribution  
of material with temperature in the wind 
which can affect both the ionization equilibrium and the
formation of the lines. 

Variation of the \ion{C}{2} and \ion{C}{4} emission on shorter
time scales can be found in Figure 14 from HST/STIS
spectra. Substantial
changes are apparent on a time scale of years between the HST
measurements, and also over a time scale of 6 days which separates the
2010 Jan 29 and 2010 Feb 4 observations of \ion{C}{4}. The
\ion{C}{4} emission appears to strengthen with \ion{C}{2}, however the 
4 simultaneous measurements show scatter. The \ion{C}{2} line is not
broadened as much as the \ion{C}{4} emission, suggesting that the
contribution from the turbulent post-shock cooling region is less.
The positive velocity side of the \ion{C}{2} emission (\gla1335.7)
resembles the positive velocity side of the H$\delta$ line (cf. Figure
7) indicating that both may be formed in a region distinct from the
higher temperature species (\ion{C}{3}, \ion{C}{4}, etc.).

A closer look at the variations in the high temperature emissions
is provided by the sequence of individual segments of the 
\fuse\ \ion{O}{6} spectra. Under  collisionally-dominated equilibrium 
conditions, \ion{O}{6} is formed at a temperature
of 3$\times$10$^5$K.  However, the \ion{O}{6} line continues to be
formed in hotter coronal plasmas due to the extended wing of Li-like
ions in the ionization balance at high temperatures. \ion{O}{6}  has been used consistently as a
diagnostic of the dynamical properties of outflows in the solar corona 
(Kohl \etal\ 2006) and X-ray emitting shocks in OB-star winds (Lehner
\etal\ 2003).
Evidence of the dramatic changes in emission from TW Hya  can be found from 
the individual \ion{C}{3} and \ion{O}{6} profiles assembled over a  32 hour span and
shown in Figure 15.  Each segment represents an integration time 
ranging from 2600 to 3700 s.  The impression that \ion{C}{3} varies more
than \ion{O}{6} is quantitatively confirmed by an extraction of the line
fluxes. \ion{C}{3} varies by a factor of $\sim$1.8 and \ion{O}{6} by a 
factor of $\sim$1.5. Evaluating the negative and positive velocity sides
of each line shows that the fluxes are not correlated between the sides. 
However the uncertainty in flux on the negative velocity side is
larger because the total counts are lower.  And the total counts in
the \ion{O}{6} line are a factor of 4 or more than in the \ion{C}{3} line, 
making the \ion{O}{6} parameters a more secure measurement.  
The relatively constant profile of the outflowing wind
absorption at negative velocities contrasts dramatically with 
the changing emission at positive velocities during this
time.   Such  profile variation  provides additional evidence that the 
emission  arises from the turbulent post-shock accretion volume, and the
the line is scattered in the hot wind which is  present 
during these observations.  Similar behavior in \ion{He}{1} 5876\AA\
was found in another accreting T Tauri star, RU Lupi (Gahm \etal\ 2013).

Fluxes taken from  the individual spectra are shown in Figure 16 where
in the first pointing the \ion{C}{3} (\gla977) line is stronger than
\ion{O}{6} (\gla1031.91); this behavior is reversed in the second
pointing about one day later.  Here we have selected segments 
where the SiC detector (containing the \ion{C}{3} line) is
well-aligned with the LiF detector (containing the \ion{O}{6} line).{\footnote The FUSE
telescope contained 4 coaligned optical channels.  Pointing on a target was
maintained by the fine error sensor which viewed one of the channels. Because
thermal changes on orbit caused rotation of the mirrors, this could lead to
misalignment of the optical channels and a target could drift in and out of the
aperture in the channels not used for guiding.  This could affect the 
measurement of the \ion{C}{3} line since it occurs in a channel not
used for guiding.   In our analysis, segments for
\ion{C}{3} were selected when the channels were coaligned.}    
However, the \ion{C}{3} atom has a metastable level
whose population is dependent on density which could contribute
to the behavior of \ion{C}{3}. And of course, the post-shock cooling
region can change as well.   During the second pointing when both lines
could be measured simultaneously over $\sim$5 hours, the fluxes are
correlated with a linear correlation coefficient of 0.9.

\section{Remarks concerning a Hot Wind}

The presence of a hot wind from TW Hya has been questioned by
Johns-Krull and Herczeg (2007), hereafter JKH.  These authors offer
several arguments purporting to show that the wind is not hot. We
argue below why we believe  that their concerns are not valid, and discuss four relevant
issues.

{\bf (1)} As evidence for outflow, JKH require sub-continuum 
absorption in a line profile to be present at `blue-shifted' 
wavelengths.  While that phenomenon is well-understood from
characteristic P Cygni line profiles,  
the signature of outflowing material can be revealed in
other ways.  A shift of the apparent wavelength of
the centroid of an emission or absorption line can reveal mass motions. 
Mass motions can also be signaled by line
asymmetries.  Hummer and Rybicki (1968) first pointed out
that apparently redshifted line profiles can arise in a differentially
{\it expanding} atmosphere; in fact these have been widely observed 
in luminous stars (cf. Mallik 1986, Robinson \& Carpenter 1995, Dupree \etal\ 2005b,
Lobel \& Dupree 2001) and in the Sun (Rutten \& Uitenbroek 1991).
The assertion by JKH that sub-continuum
absorption must be present to identify outflow neglects both line-transfer
physics and stellar observations themselves. 

{\bf (2)} In an attempt to demonstrate the absence of absorption
in resonance doublets, JKH arbitrarily
scale and overlay the long wavelength member of the \ion{C}{4}
doublet on to the short wavelength member of the doublet by forcing 
both the line peaks and blue wings to match. They claim that signs of 
absorption are missing in the red wing of the 1548\AA\ line  
that would be caused by the wind absorption in the 1550\AA\ line. Such
arbitrary scaling is not  justified.  Atomic
physics, supported by  laboratory measurements,  specifies the wavelength 
separation between these lines as well as the ratio of the line oscillator
strengths.  In an effectively optically thin doublet
such as \ion{C}{4}, the line flux ratio (\gla1548/\gla1550) is 2. Since it is preferable
not to amplify the errors in the weaker line of the 
doublet, here we place the stronger (1548\AA) line 
onto the weaker (1550\AA) transition.  The original HST data used
by JKH, namely {\it O58D0130}, has been superseded by a 
more optimal spectrum reduction, namely CoolCAT (Ayres 2010) and we also
consider that reduction (see Table 1). In Figure 17,
we show the CoolCAT reprocessed profile as well as the original spectrum
used by JKH.\footnote{More recent papers (Yang \etal\ 2012) than
  Herczeg \etal\ (2002) estimating  the UV line fluxes from 
TW Hya using revised calibrations report a substantially 
increased value of the \ion{C}{4} flux by a 
factor of $\sim$5.}

When the \ion{C}{4}
1548\AA\ line is  properly overlaid on the 1550\AA\ line 
by shifting it according to the wavelength separation (2.577\AA,
Griesmann \& Kling 2000), and scaling the flux of 1548\AA\ by 2, 
two facts are immediately apparent from either data reduction: 
(1) the wind opacity in the resonance line 1548\AA, indicated by the negative velocity
side of the profile, is greater than in the subordinate line 1550\AA,  as expected from
atomic physics, and  (2) excess
wind absorption does indeed occur in the long wavelength wing of 1548\AA\ caused by wind absorption
from the 1550\AA\ line.  Consideration of the errors in the measured fluxes at individual
wavelengths in the HST/STIS spectrum  indicates that  the separation 
visible in Figure 17 between the 1550\AA\ line and the 
shifted scaled 1548\AA\ line ranges from $\sim$2.5 to 3.2$\sigma$ in the velocity region 
from $+$175 to $+$350 \kms. Moreover, in  this region the dashed line in Figure 17 systematically  
lies below the solid curve. Thus the effects of hot wind absorption 
are clearly evident in the \ion{C}{4} profiles.

{\bf (3)} Trying to demonstrate the absence of wind absorption
by the \ion{O}{6} 1037.61\AA\ transition affecting the \ion{C}{2}
1037.02\AA\ line, JKH  resort to an {\it ad hoc} construction.  In
this case, 
both the activity  
of TW Hya and atomic physics  undermine their arguments.  The
\ion{O}{6}  1037.61\AA\ 
line, has a neighboring \ion{C}{2} line separated by  $-$172  \kms\ which
places the \ion{C}{2} line in the velocity range of scattering produced by a hot
\ion{O}{6} wind. To attempt to show  that the \ion{C}{2} line is not weakened by
scattering, JKH predict the intrinsic strength of a blended \ion{O}{6} and
\ion{C}{2} feature.
By taking the \ion{C}{2} line profile at 1335.71\AA\ observed in 2000
with HST/STIS, these authors scale the profile to
mimic the \ion{C}{2} transition at 1037.02\AA, 
and add it to the line profile of \ion{O}{6} 1031.91\AA\ 
observed in 2003. The resulting scaled combination is then compared
to the blended \ion{O}{6}(1037.61\AA) and \ion{C}{2}(1037.02\AA)
emission feature. 
JKH do not present the parameters used for  this scaling, 
shifting, and combination.  

From first principles, the intrinsic 
relative strength of the \ion{C}{2} 1335.7\AA\ line to the
\ion{C}{2} 1037.02\AA\ line is hard to predict.  The ratio of $gf$\gla\ values
gives a factor of 5.3 for the $\lambda\lambda$1335.7/1037.02 ratio, and the Boltzmann factor ratio could be
2.4 at a temperature of 40000K suggesting the 1037.02\AA\ line
could be 13 times weaker than the 1335.71\AA\ line under optically thin conditions
in LTE, and not 1.5 times 
weaker as JKH  propose.  However unknown   
optical depths and non-LTE conditions occur 
in the formation of resonance lines of \ion {C}{2} causing them
to appear weaker and decrease the $\lambda\lambda$1335.7/1037.02 ratio.
These issues confound any meaningful scaling. Moreover, the line
profiles themselves  change with time as demonstrated earlier.
Comparison of  HST spectra from 2000 to  FUSE
spectra from 2003 (cf.  Figure 13) clearly shows the substantial changes
in the emission profiles of ultraviolet lines on the `red' sides 
($+$100 to $+$400 \kms) between these three years.
It is obvious from HST spectra of \ion{C}{2} in 2000, 2002, 
and 2010 (Figure 18) that the critical region
of the 1335.7\AA\ line profile used to assess absorption, namely the
region from $+$35 to $+$85 \kms, varies in both absolute flux as well as
slope.   The claim made by JKH ---  that
this scaled,  shifted,  blended feature does not exhibit strong  
absorption on the red side caused by
the hot \ion{O}{6} wind --- is unjustified, in particular given 
the  substantial changes in both the emission and the wind.

{\bf (4)} JKH suggest that two H$_2$ lines arising from the
 circumstellar
disk  exhibit behavior
that suggests a cool wind is present, but not a hot wind.  Their
interpretation does not agree with the geometry of TW Hya and its disk,
nor with the known characteristics of a wind from a dwarf star. 

The  H$_2$ line at 1333.85\AA\ occurs  
$-$165 \kms\ from the \ion{C}{2} line at 1334.5\AA\ (at $-$430 \kms\
in Figure 18).  
JKH claim that this  H$_2$ feature is absorbed by the \ion{C}{2} wind because 
their model overpredicts its strength in their spectrum taken
in 2000. By contrast, an H$_2$ line lying at a velocity of $-$165
\kms\ from the \ion{C}{4} 1548\AA\ feature does not suffer wind absorption
when compared to their model (Herczeg \etal\ 2002).  Thus JKH conclude
that a cool wind  exists, but not a hot wind. 

Many spectra subsequent 
to the one used by JKH reveal conflicting patterns (see Fig. 18).
In 2002, the \ion{C}{2} wind extends to $-$251 \kms, which would
weaken not only the line located at 1333.85\AA, but should
weaken another H$_2$ line located at $-$225 \kms\  
(corresponding to $-$500 \kms\ in Figure 17), but 
it does not.  Additionally, in the presence of a strong
\ion{C}{2} wind,  another H$_2$ line emerges 
at 1335.2\AA\ ($\sim -$125 \kms\ in Fig. 18; the R(2) 0-4 transition)
which did not appear previously.  All of
these spectra indicate optically thick winds in \ion{C}{2} are
present and the wind opacity is variable.  Furthermore,  the  
H$_2$ lines behave in apparently unpredictable fashion -- perhaps
because they are produced  through photoexcitation by  variable
Lyman-\gal\ emission. With a more precise
location of the  H$_2$ emission, the behavior of
the molecular lines might aid in the wind description.  However,
since these lines are pumped by H-Ly\gal, which itself varies,
trying to pin down a consistent model may prove challenging. 
Moreover, serious considerations discussed earlier in (3), rule out the putative effects
of wind absorption.

Trying to assess potential effects of a wind upon the observed 
H$_2$ emission involves many factors. First, the line opacity 
in the wind must coincide in wavelength with the H$_2$ emission 
from the disk.  H$_2$ is believed to be produced
less than 2 AU from the star (Herczeg \etal\ 2002) where it 
originates in the gaseous component of
the circumstellar disk around TW Hya.  
Recent ALMA observations suggest the dust-depleted 
cavity lies within 4 AU of 
the star (Rosenfeld \etal\ 2012).  Ions in a radially-moving stellar 
wind passing over a distant disk will
absorb at lower velocities than when observed against the 
nearly pole-on  star itself. 
Rosenfeld \etal\ (2012) suggest the TW Hya disk might have
a warp of $\sim$8 degrees, such that the effective velocity of any
absorption would occur at a still lower velocity.
Thus the wind may not have significant opacity at the position
of the H$_2$ emission when passing across the disk at a low angle.  Moreover, 
any channeling, flux-tube expansion,  or inhomogeneities (all of which
occur in the solar wind)  will modify the amount of absorption
and the velocity at which it occurs.  A more fundamental concern
appears to  be the ionization stage in the wind itself.    If 
the stellar wind at 2 AU does not contain \ion{C}{2} or \ion{C}{4},
then absorption will not occur.  From very detailed knowledge about the wind of one star,
our Sun, the low species ions are effectively absent at a distance of 
1 AU (Landi \etal\ 2012) and the wind contains the ionization stages of the 
corona. Thus at a distance of 1 AU or larger, the wind of TW Hya, if similar
to the solar wind,  would appear to
lack the ions that JKH invoke for  their arguments.

The presence of a hot wind from TW Hya is not unexpected. The Sun is a low mass
star, with weak magnetic fields and it possesses an outer atmosphere with temperatures
of a few MK. Wave-heating mechanisms (Kohl \etal\ 2006) or
possibly nano-flares resulting from reconnection (Tripathi \etal\ 2010) can
heat the corona.  Self-consistent models for hot winds have been constructed
for accreting T Tauri stars (Cranmer 2008).  These winds are turbulence
driven, and the turbulence has two sources: interior convection and `ripples' 
from nearby accretion shocks.  Both observations and theory offer much
evidence that low-mass cool stars can support a hot outer atmosphere.

We conclude that the stellar wind of TW Hya is a hot wind, evidenced 
by the ultraviolet spectra as originally suggested (Dupree \etal\ 2005a).

\section{Discussion and Conclusions}

The many spectra of TW Hya  reported here document  the source and components
of emission line features and the  variability of the
accretion process,  subsequent atmospheric heating,  and wind
expansion.  We find:

1. Time-domain spectroscopy reveals symmetric H\gal\ profiles in approximately 
half of our spectra and indications of outflowing wind asymmetries in the remaining
spectra. This spectroscopic example  indicates that the source of the
broad emission feature arises from the  TW Hya star itself, and  can 
be identified with the post-shock cooling zone.
Broad lines ranging from X-ray through
infrared transitions can be  straightforwardly explained by the
presence of a variable  turbulent post-shock cooling volume producing the
emission  which can be
modified by wind scattering and/or material infall. 
Earlier evidence of a logical progression in line shape
and flux following an X-Ray accretion event also supports this 
interpretation (Dupree \etal\ 2012).
Models are now needed for the temperature structure and emergent
emission from this post-shock region.  A recent MHD simulation
(Matsakos \etal\ 2013) of a two-dimensional accretion shock impacting a
chromosphere suggests that substantial chaotic motions can result in
the low chromosphere.  These need to be followed up with calculation
of the radiation emitted from the cooling plasma.

2. Absorption features at negative velocities  in the H\gal\ line
 noted by Alencar et al (2002), are demonstrated with the frequent continuous
spectral coverage  reported in this paper,  to be stable in velocity during a night's
observation. Moreover, their stable behavior  and opacity characteristics 
suggest they appear to arise from the 
silhouette of an accretion stream high in the stellar chromosphere. 
At positive velocities, absorption
in H\gal\  mimics the infall speeds detected in \ion{He}{1}.

3. Current models (Muzerolle \etal\ 2001; Kurosawa \& Romanova 2012, 2013)  
 that consider the  optical  emission from 
accretion `funnels' do not agree with the observed line
profiles in TW Hya.  The observed profiles lack the shift, the
asymmetries and/or absorption predicted from accretion stream
models. In addition, many models do not
include a stellar wind, which is clearly indicated in the
observed line profiles, nor emission from a post-shock cooling region.

4. A hot wind, and its variable speed and opacity are  
revealed by the UV and far-UV resonance lines with spectra spanning 
a decade in time. Previous arguments offered (JKH) against the presence of  
a hot wind are shown to be unpersuasive. Broad line profiles and
variability support  both  formation in a turbulent
medium and the presence of a hot wind.   This hot wind is a natural
extension of the wind indicated by the near-IR helium line profile.

5. The \ion{He}{1} near-IR transition at 10830\AA\ shows
less night-to-night absorption variability than H\gal, suggesting a large-scale 
wind outflow from the star. The emission in the helium line correlates inversely 
with the velocity of the accreting inflowing material 
consistent with a simple model of dipole accretion, and formation of
the
emission in the turbulent post-shock cooling zone.  The stellar wind
velocity appears related to the strength of helium emission providing
direct spectroscopic evidence  that the 
accretion process leads to wind acceleration.

\bigskip

\acknowledgments
We thank the anonymous referee for his/her comments that improved
the manuscript. 
The authors gratefully acknowledge the helpful support from
Gemini-S astronomers, the staff at Magellan, and KECK II
while acquiring these spectra. Observers at FLWO were a
great help on short notice.  This research has
made use of NASA's Astrophysics Data System Bibliographic Services.
Some of the data presented here was obtained from the Mikulski
Archive for Space Telescopes (MAST). STScI is operated by the Association
of Universities for Research in Astronomy, Inc. under NASA contract
NAS5-26555. Support for MAST for non-HST data is provided by the NASA
Office of Space Science via grant NNX09AF08G and by other
grants and contracts.   We wish to extend special thanks to those of
Hawaiian ancestry from whose sacred mountain of Mauna Kea we are 
privileged to conduct observations.  Without their generous 
hospitality, the Keck results presented
in this paper would not have been possible. 

\bigskip 

{\it Facilities:} \facility{Gemini:South
  (PHOENIX)}, \facility{Magellan:Clay (MIKE), \facility{HST(STIS)},
\facility{FUSE}}, \facility{FLWO:1.5m (TRES)}, \facility{Keck II (NIRSPEC)}


\begin{figure}

\hspace*{-0.3in}
\includegraphics[angle=90.,scale=0.3]{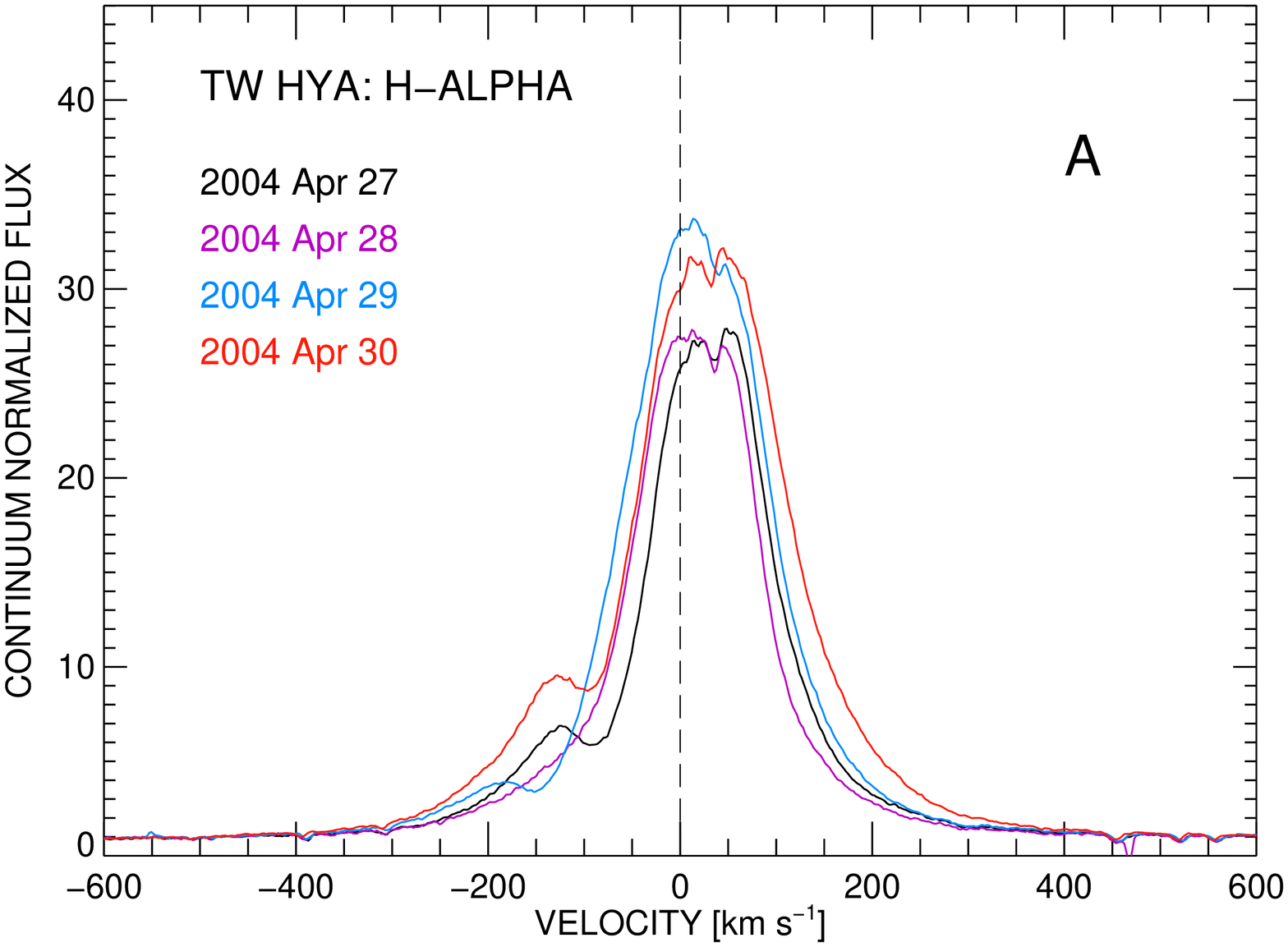}

\vspace*{-2.1in}

\hspace*{3.0in}
\includegraphics[angle=90.,scale=0.3]{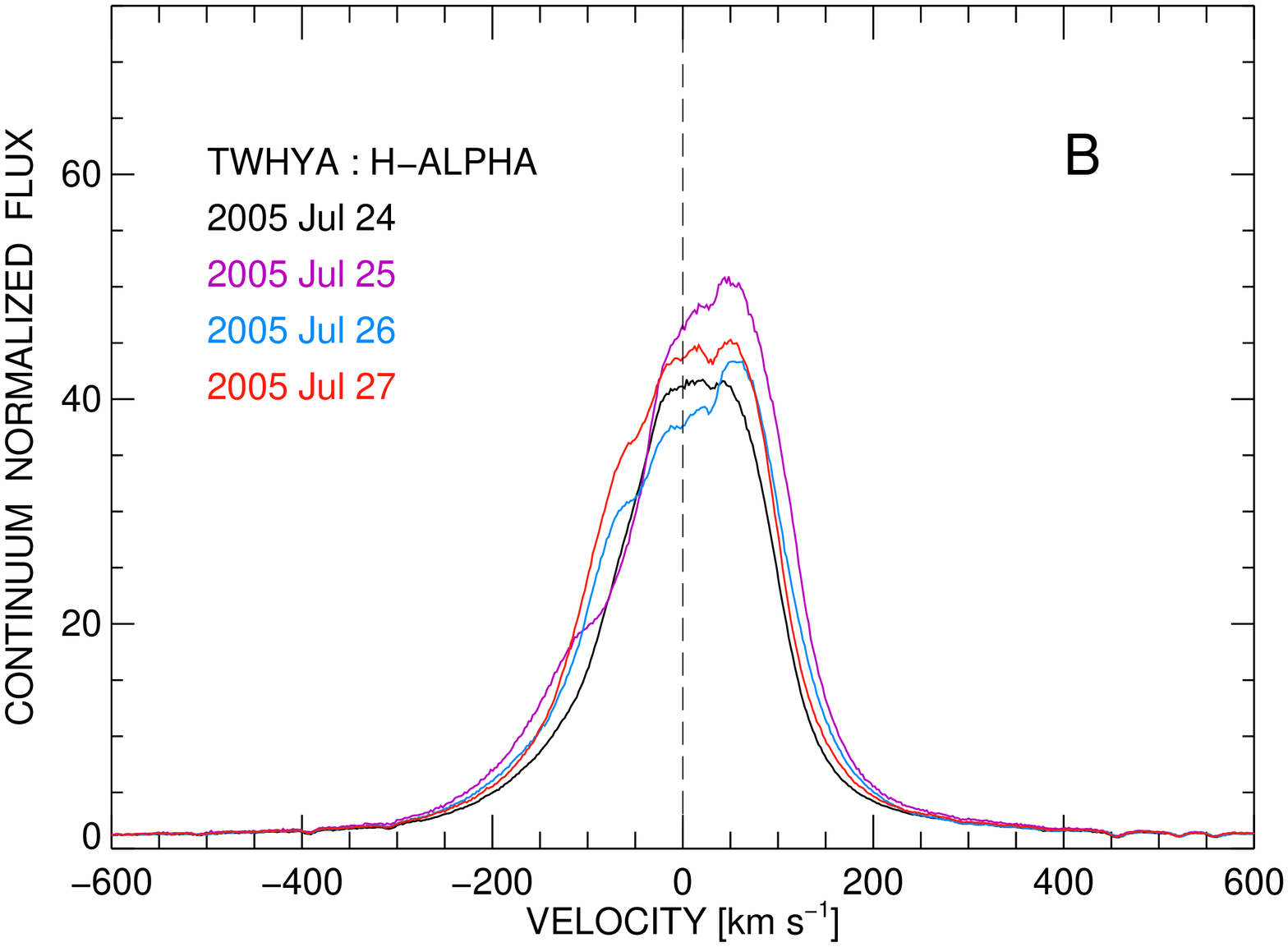}

\vspace*{+0.1in}
\includegraphics[angle=90.,scale=0.3]{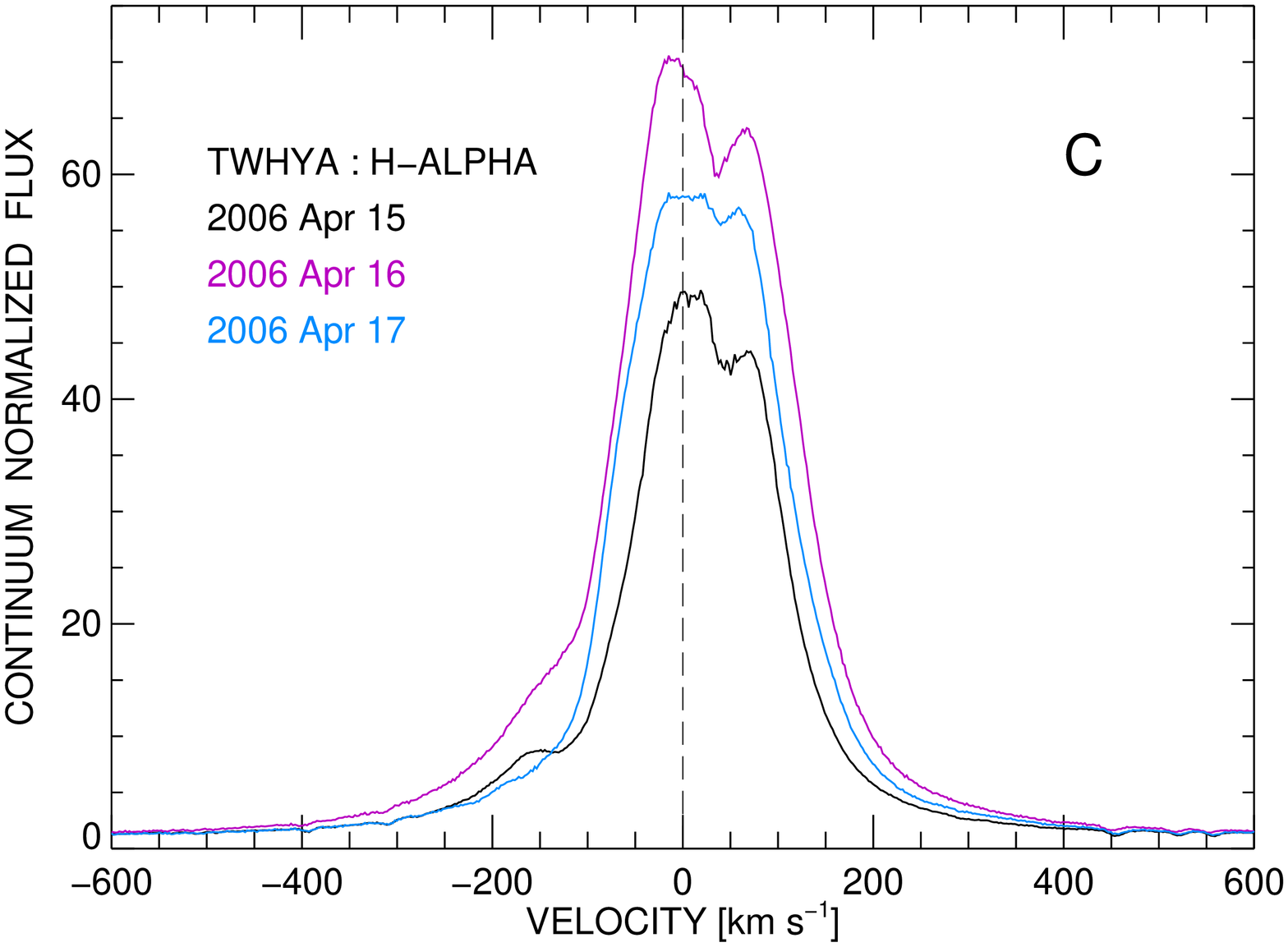}

\vspace*{-2.1in}

\hspace*{3.0in}
\includegraphics[angle=90.,scale=0.3]{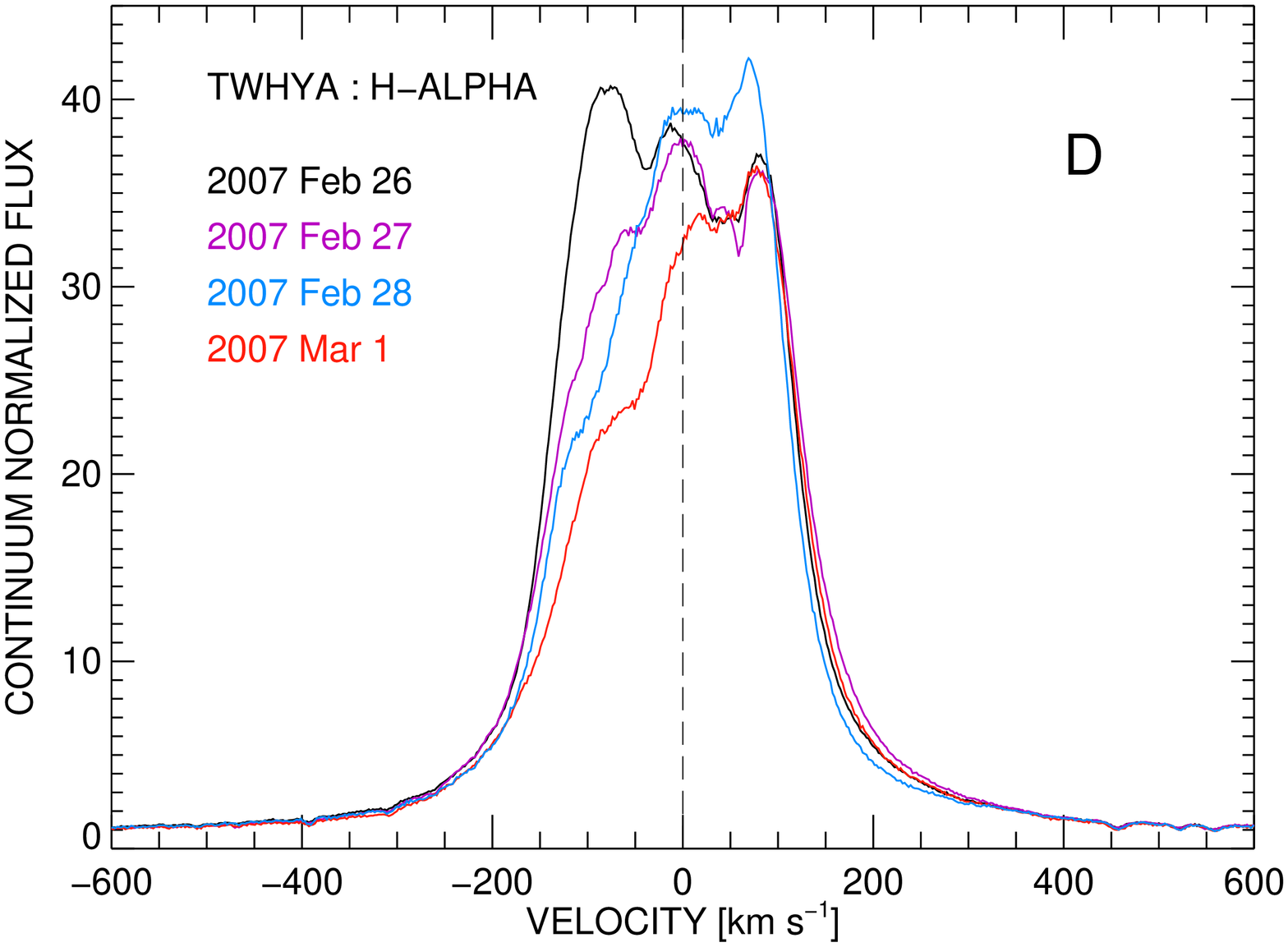}

\vspace*{+0.1in}
\includegraphics[angle=90,scale=0.3]{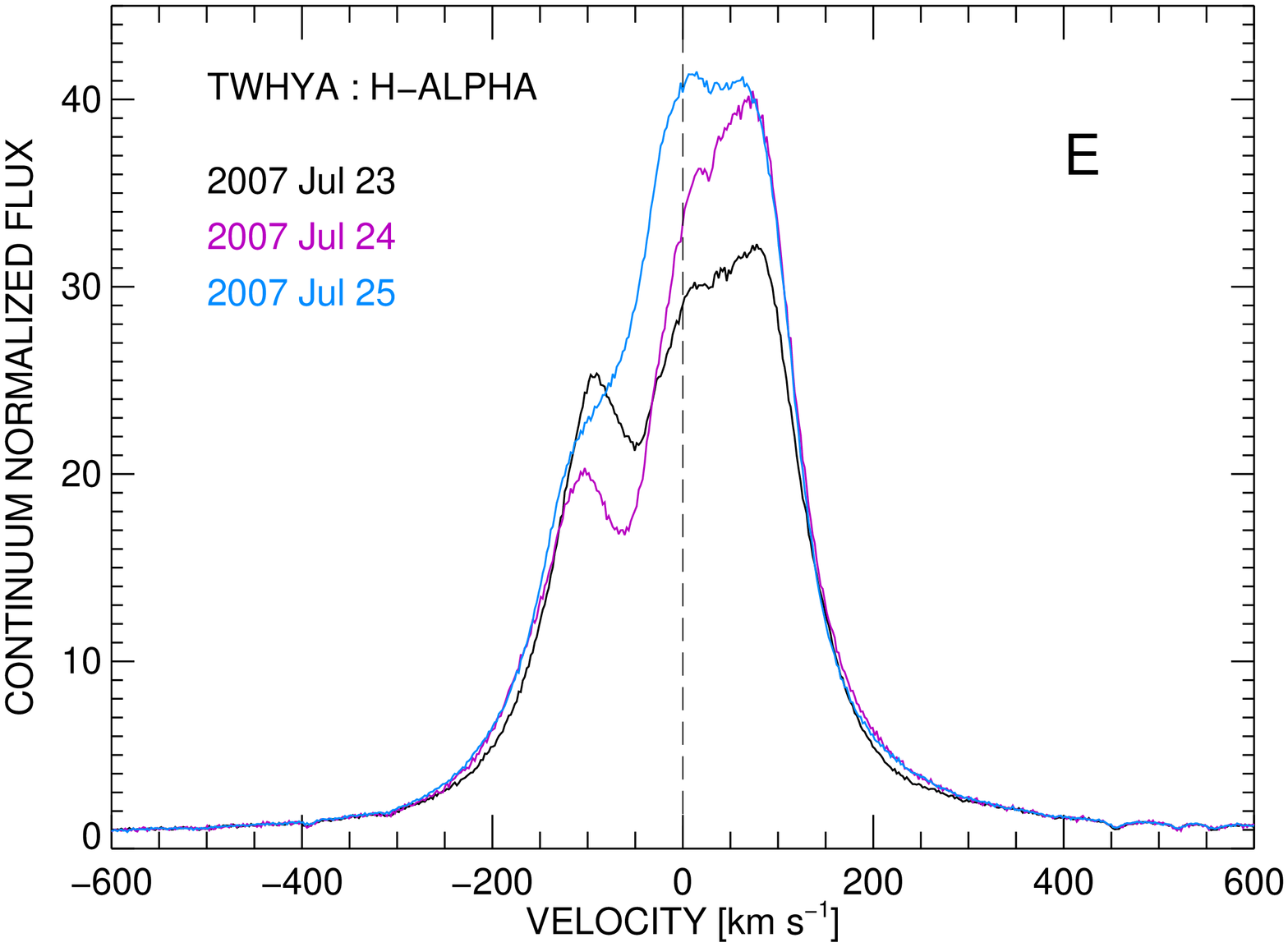}

\vspace*{-2.1in}

\hspace*{3.0in}
\includegraphics[angle=90,scale=0.3]{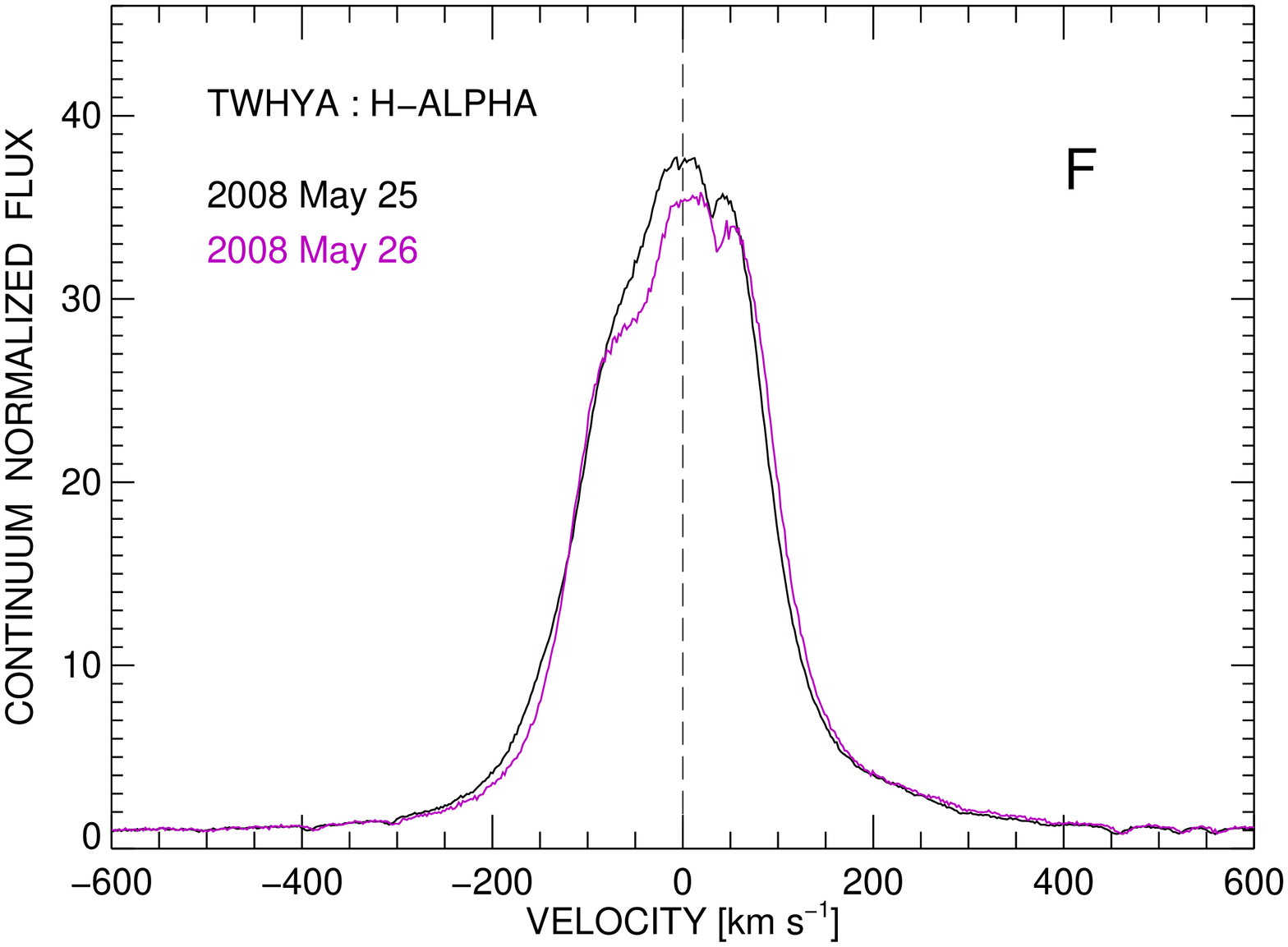}

\end{figure}


\begin{figure}
\includegraphics[angle=90.,scale=0.3]{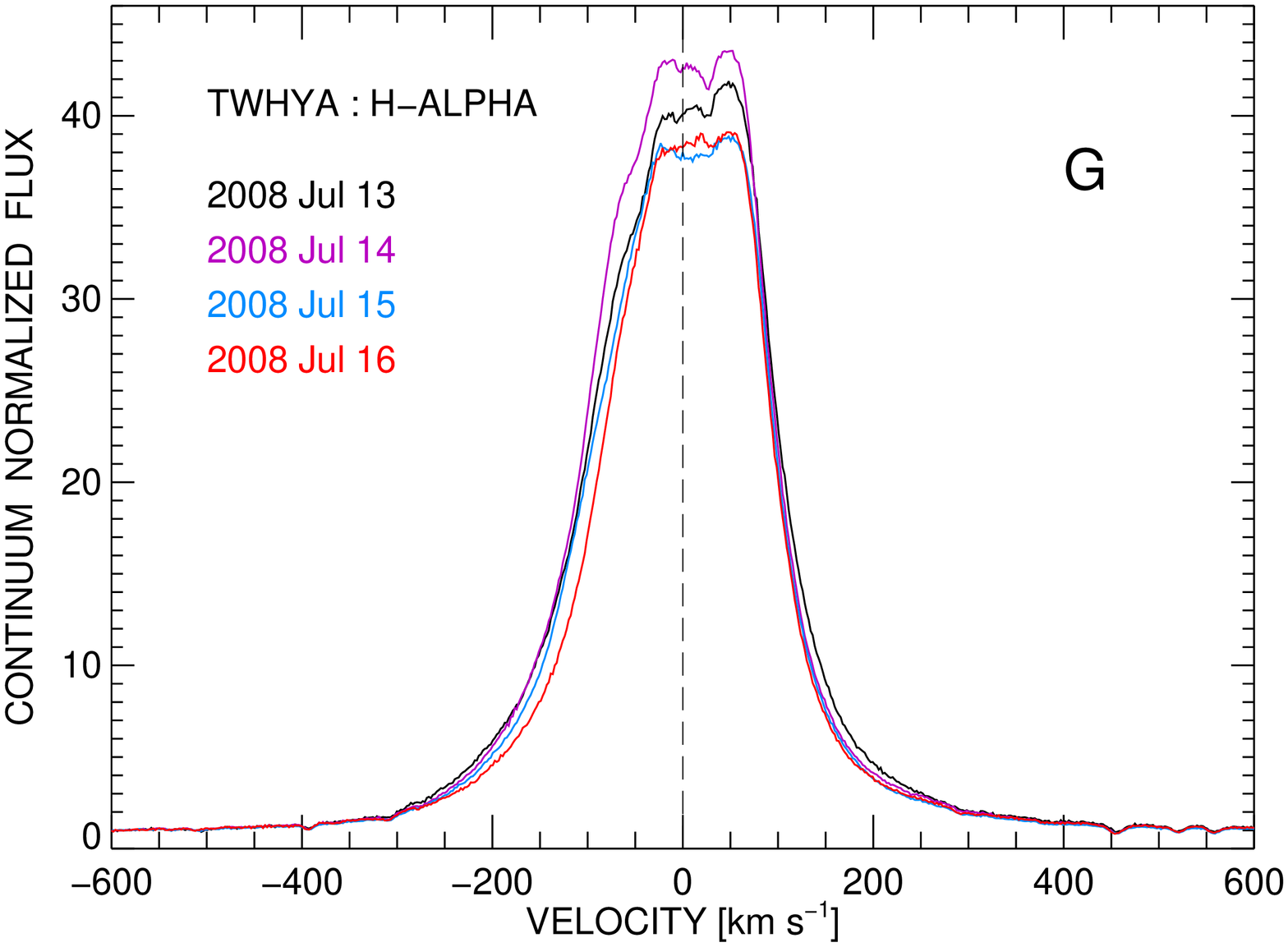}

\vspace*{-2.1in}

\hspace*{3.0in}
\includegraphics[angle=90.,
  scale=0.3]{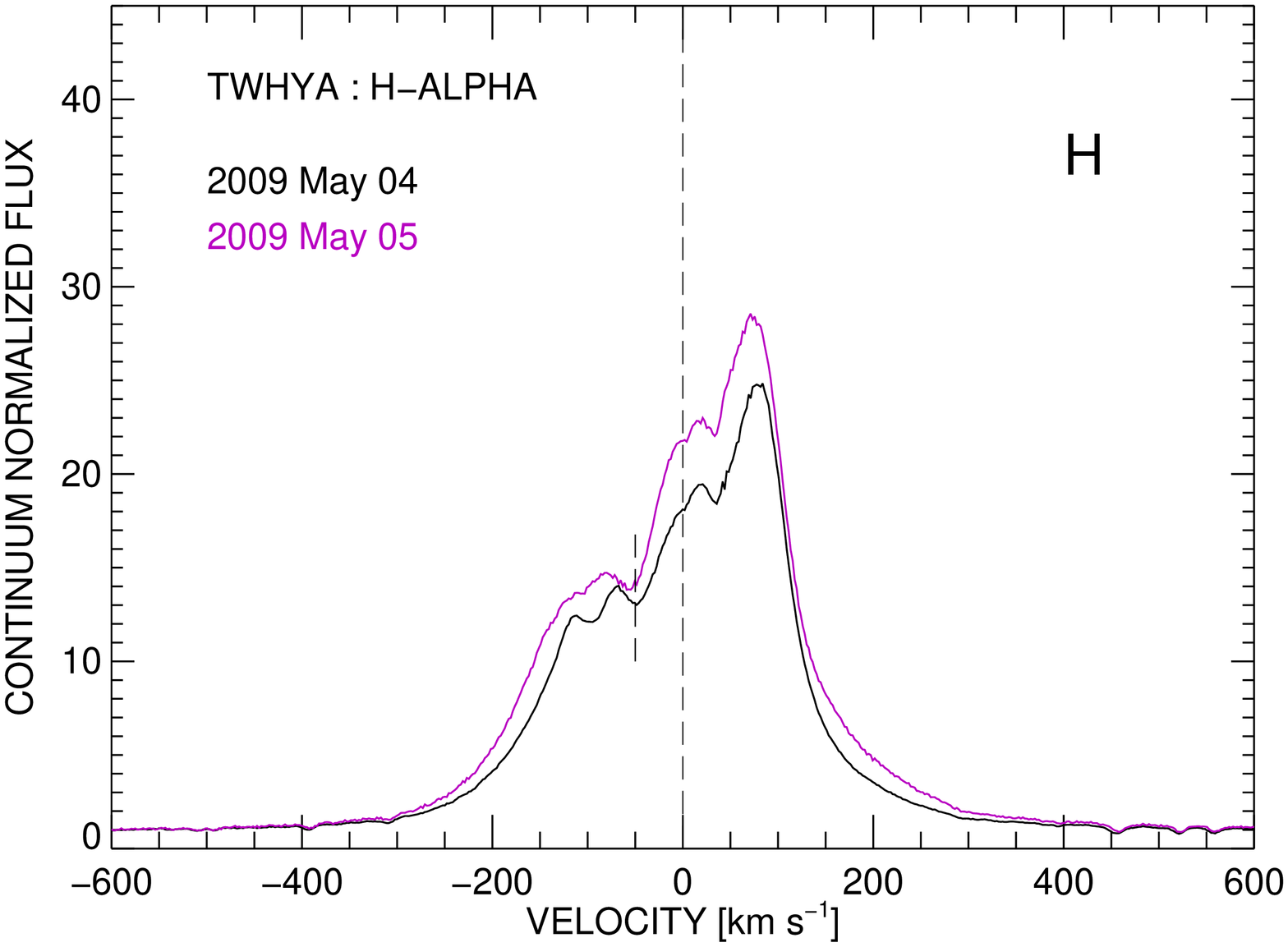}

\vspace*{+0.1in}
\includegraphics[angle=90,
  scale=0.3]{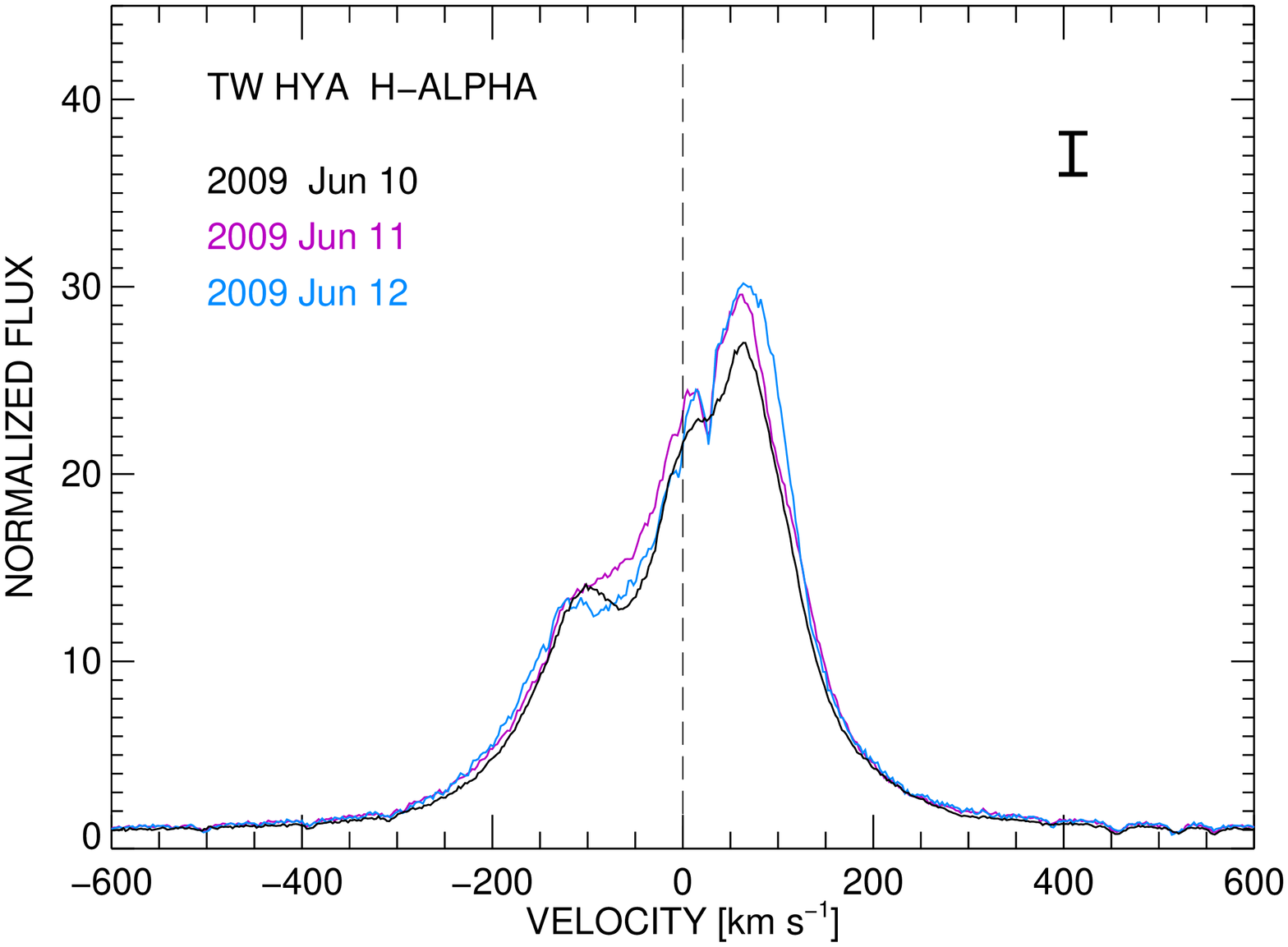}

\vspace*{-2.1in}

\hspace*{3.0in}
\includegraphics[angle=90.,
  scale=0.3]{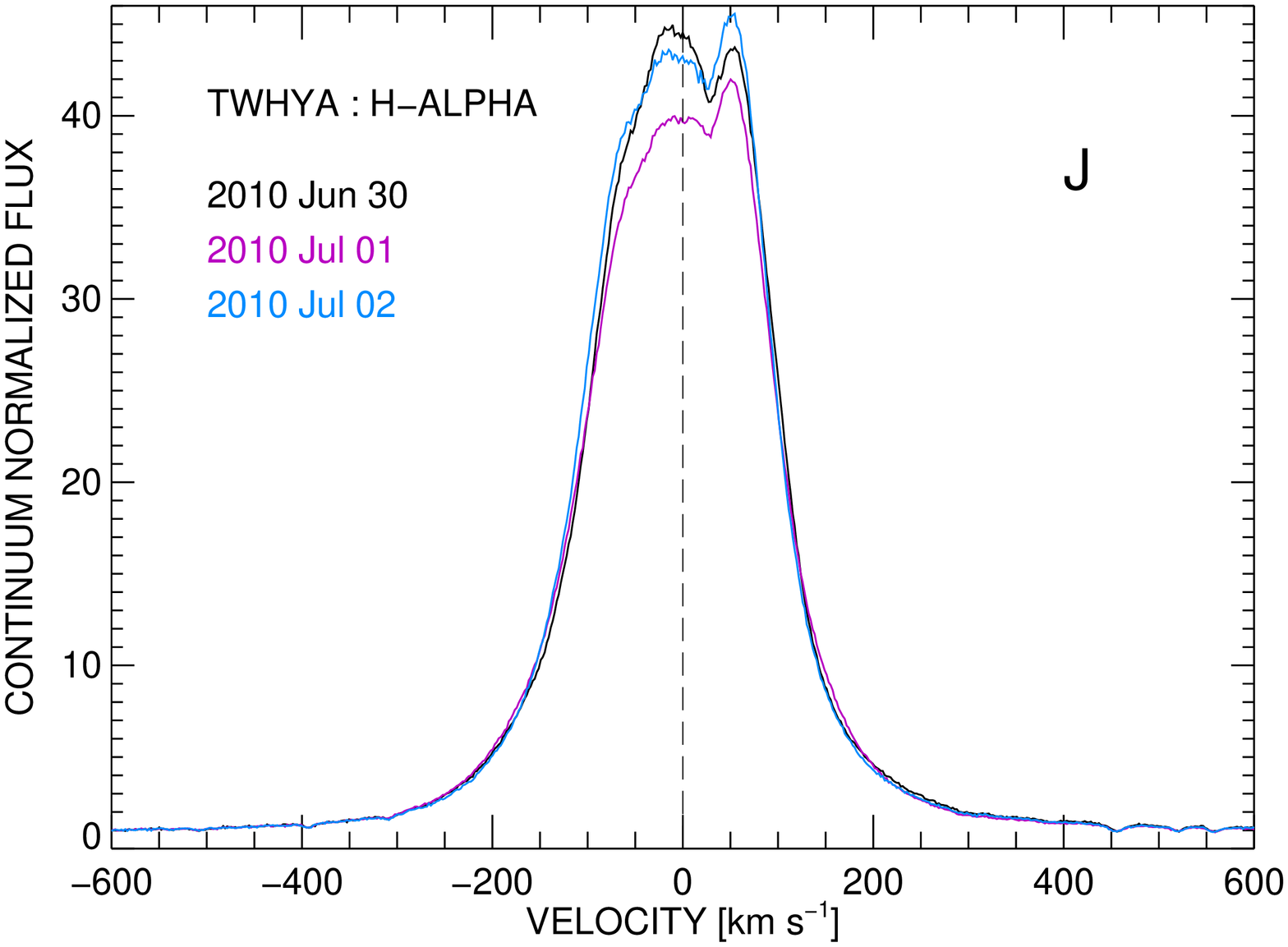}

\vspace*{0.1in}
\includegraphics[angle=90.,scale=0.3]{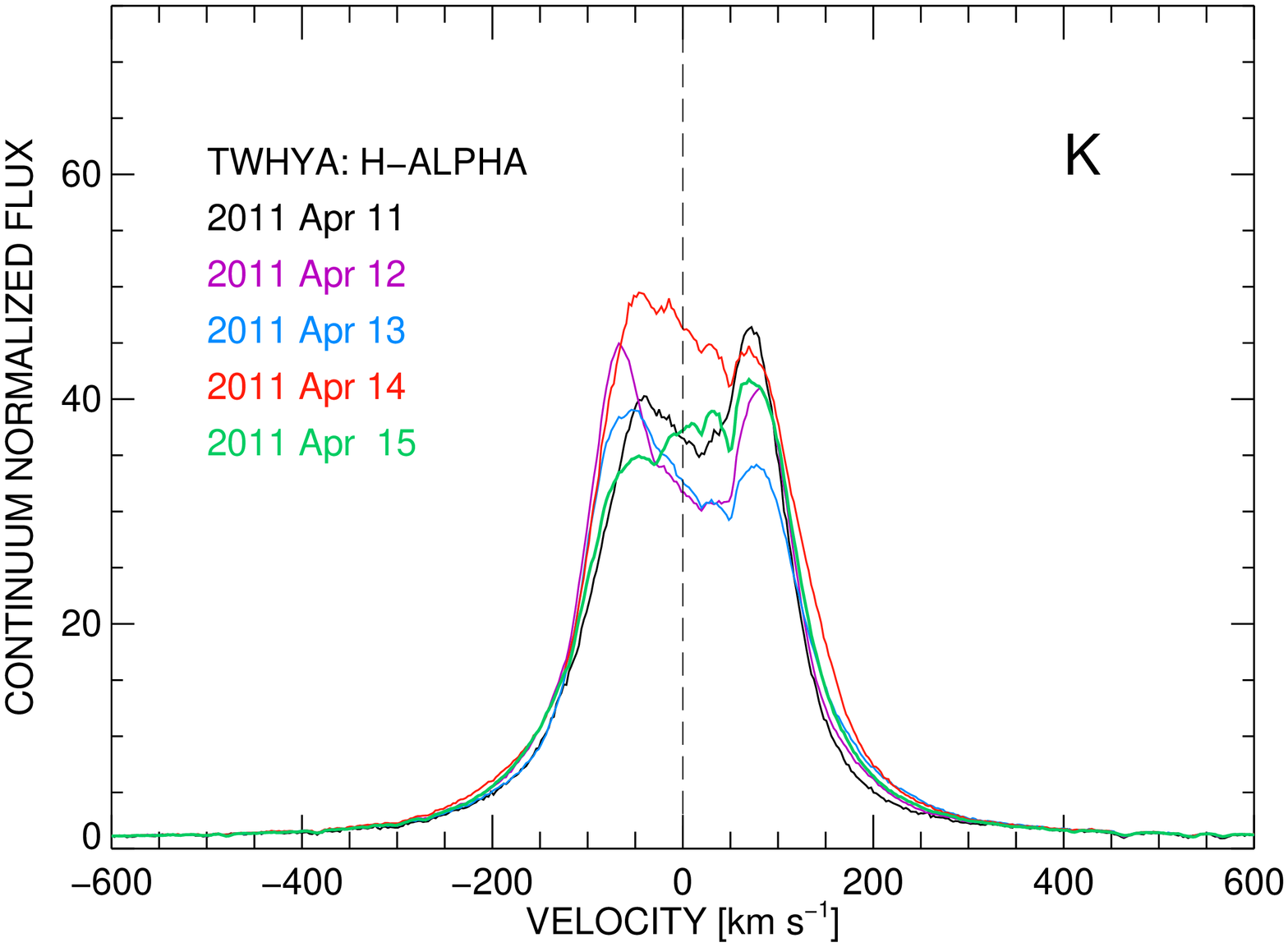}

\vspace*{-2.1in}

\hspace*{3.0in}
\includegraphics[angle=90.,
  scale=0.3]{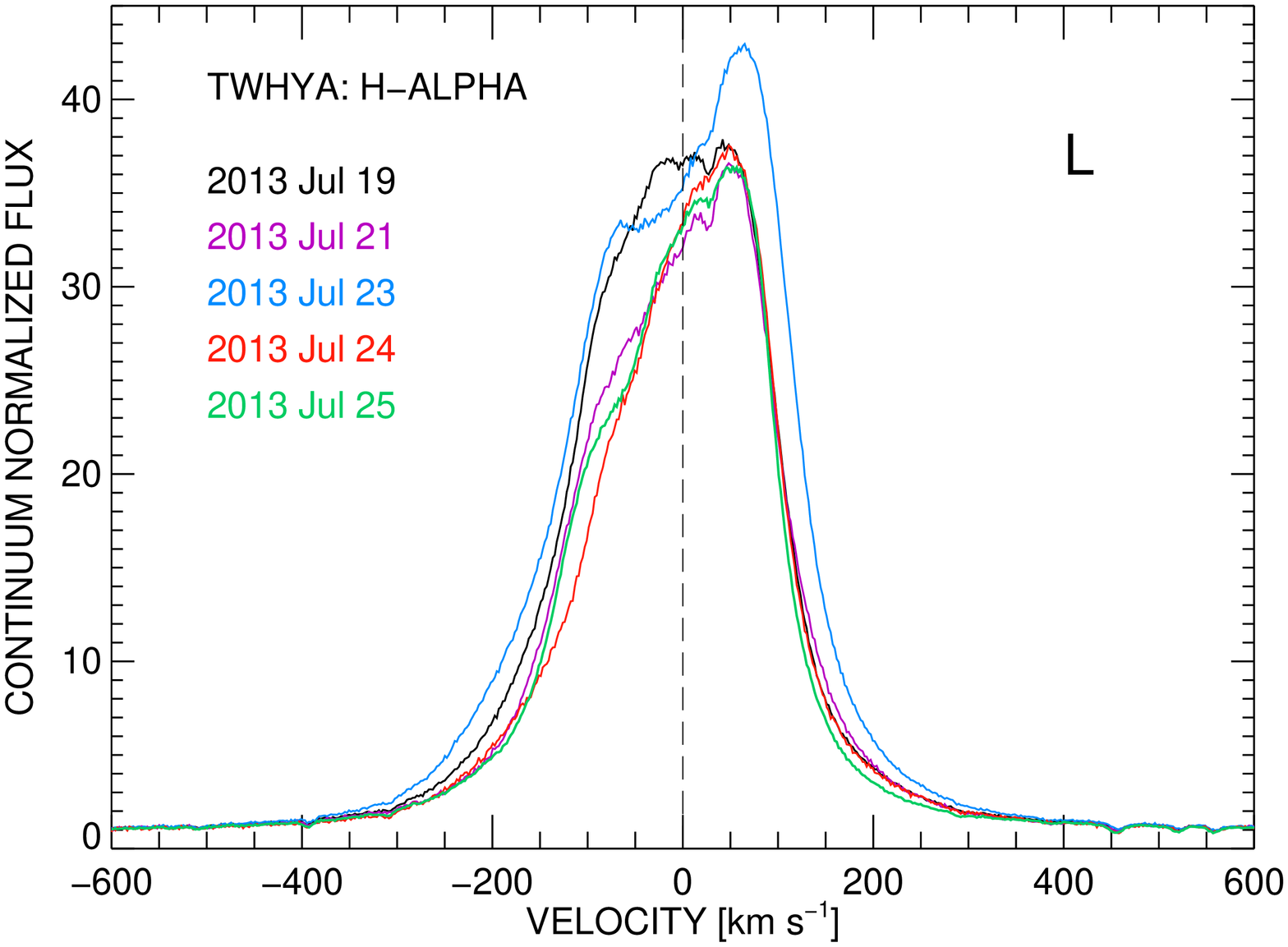}

\caption{Single H\gal\ spectra of TW Hya selected from  ten years of 
measurements from 2004 to 2013. Note
  that there are two different scales on the y-axis.
  Observations in 2005 Jul,
  2006 Apr, and 2011 Apr indicate stronger H\gal.  The narrow
  absorption feature near $+$50 \kms\ is due to water vapor. All of
  these spectra
suggest that H\gal\ is centered on the stellar radial velocity, and
  the
profile is substantially modified by wind absorption causing the apparent `red'
enhancement.}
\end{figure}

\begin{figure}
\begin{center}
\includegraphics[angle=90.,scale=0.6]{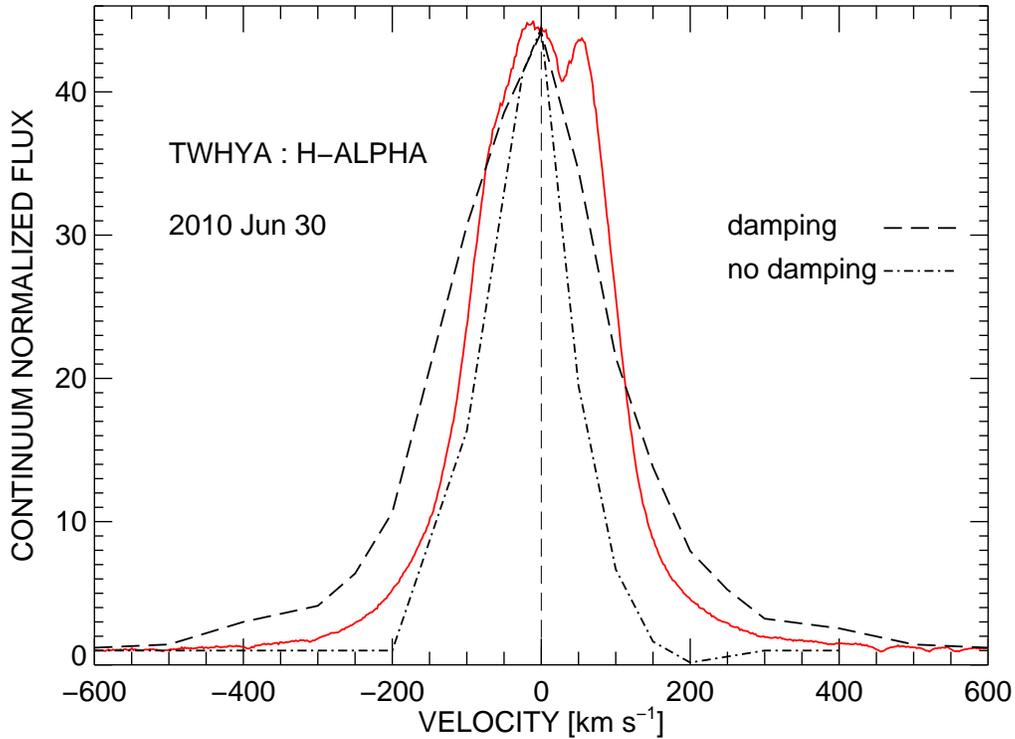}
\caption{The observed H\gal\ spectrum from 2010 Jun 30 ({\it solid line}) with two models overlaid
  from Muzzerolle \etal\ (2001) that assume either no line-damping or an
  arbitrary damping parameter.  These models assume formation of
  H\gal\ in an accretion stream which can be seen by the weaker
  positive velocity side of the profile as compared with the negative
side, and slight subcontinuum absorption
  when damping is absent.  A stellar wind is not included in the
  models accounting in part for the excess emission of the model on
  the negative velocity side.  Also the model profiles appear
  `pointed' which is not found in the observed profiles (see also Figure 1).}

\end{center}
\end{figure}

\begin{figure} 
\begin{center}
\includegraphics[angle=0.,scale=0.7]{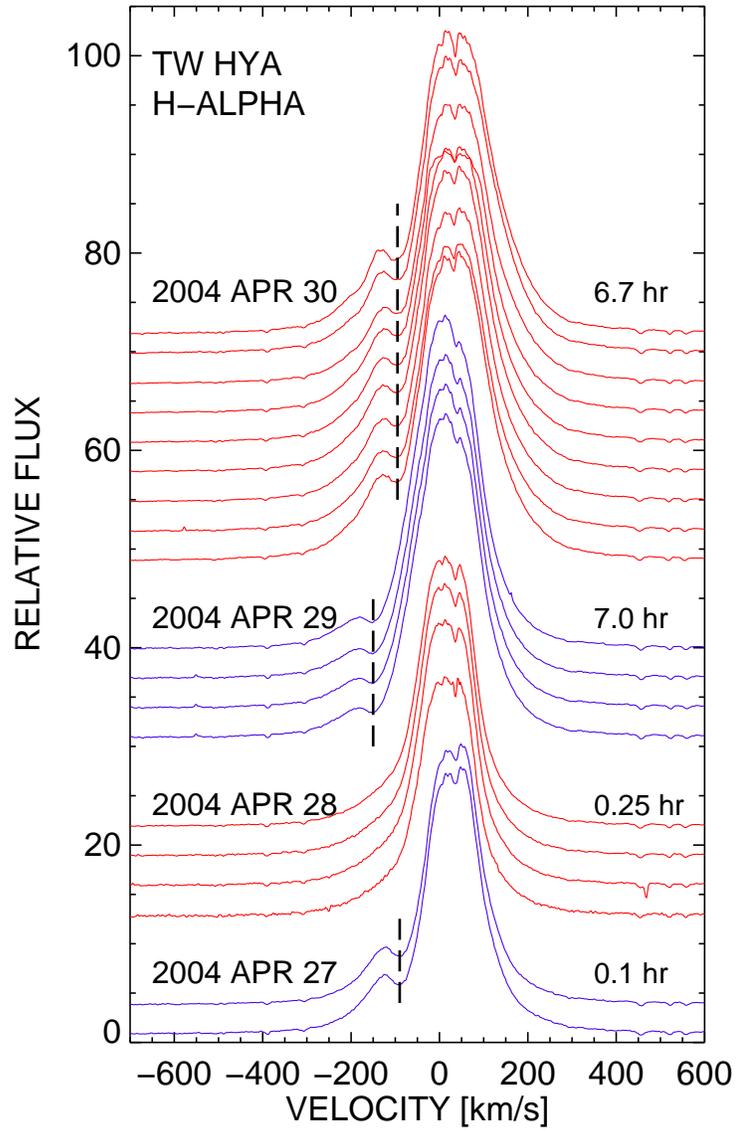}
\caption{Magellan/MIKE spectra over 4  nights in 2004 showing the appearance
  of wind structures stable in velocity. The time interval between the
  first and last spectrum is shown. Each spectrum has been offset
for display.  The broken vertical lines mark the constant velocity of
  the absorption feature during each  night. The notch in 
the H\gal\ emission at $+$50 \kms\ results 
from water vapor absorption.}
\end{center}
\end{figure}


\begin{figure}
\begin{center}
\includegraphics[angle=0.,scale=0.7]{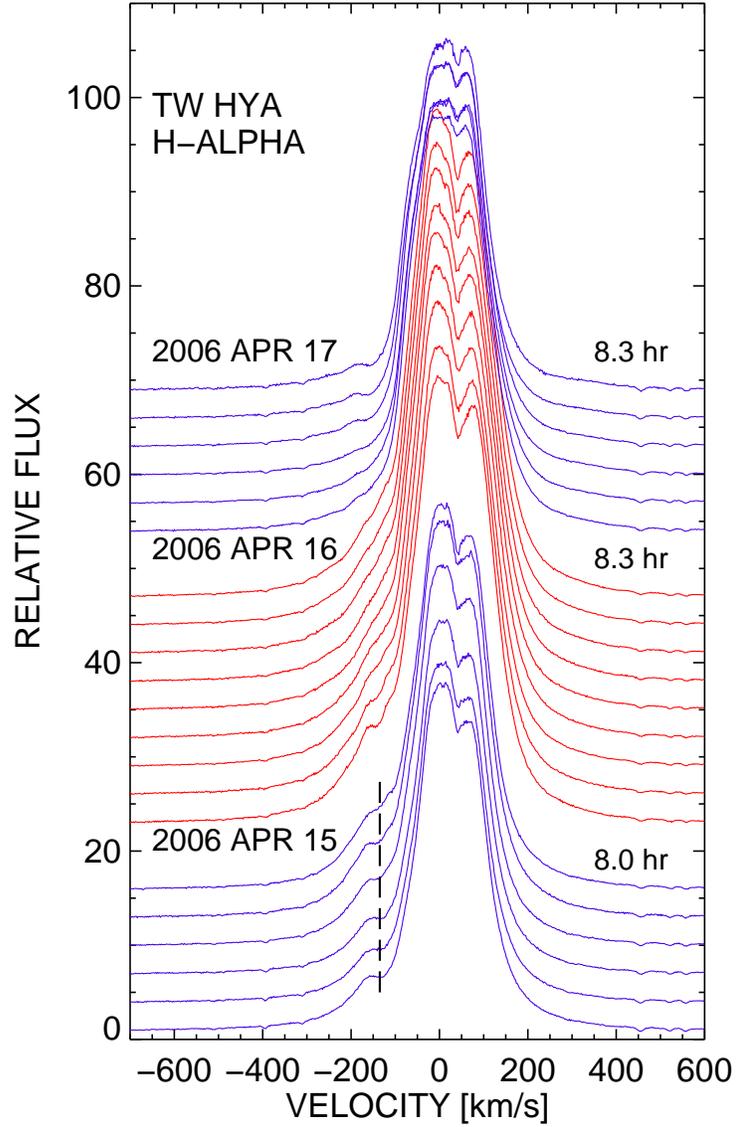}
\caption{Magellan/MIKE spectra of the H\gal\ line  over 3 consecutive nights
  in April 2006.  The narrow absorption near $+$50 \kms\ is caused by
  water vapor. More spectra  were obtained than displayed  here and a gray scale
  representation of all spectra is shown in Figure 5.}
\end{center}
\end{figure}

\begin{figure}
\hspace*{-1in}

\includegraphics[scale=0.4]{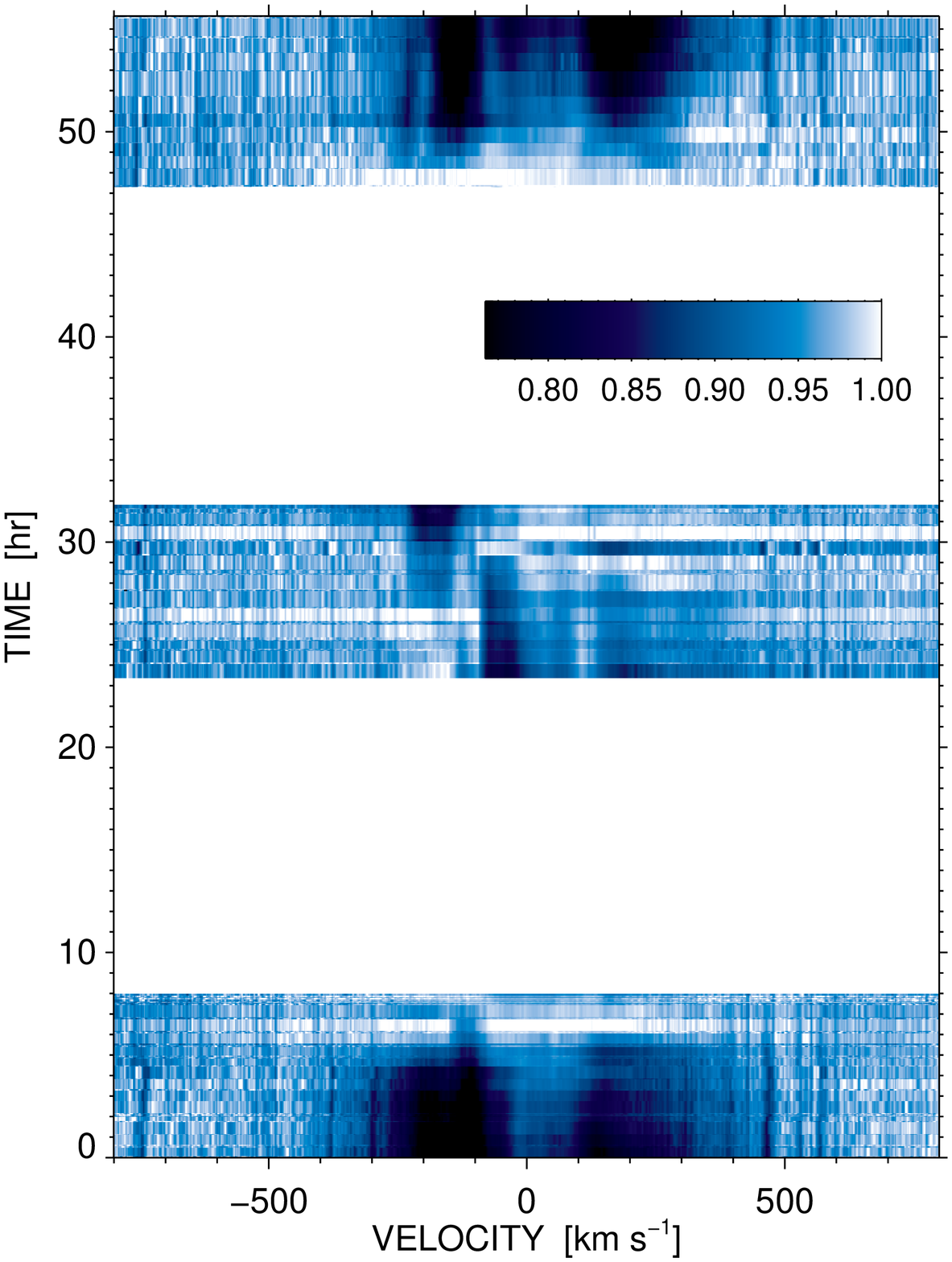}
\vspace*{-3.6in}

\hspace*{3.0in}
\includegraphics[scale=0.6]{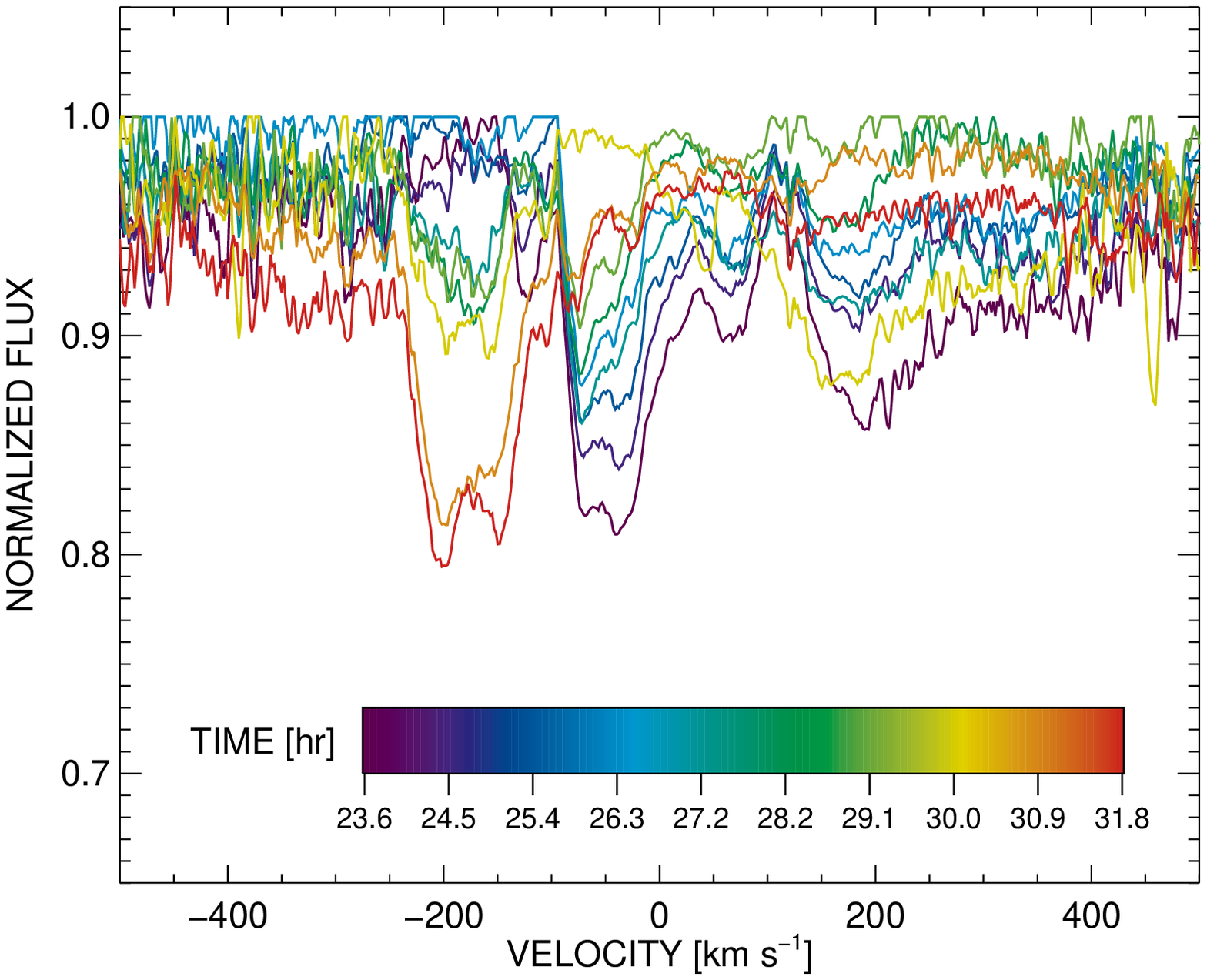}
\vspace*{0.5in}
\caption{{\it Left panel:} Gray scale representation of the H\gal\ profiles
of TW Hya over 3 consecutive nights in April 2006 where each spectrum 
has been divided by the {\it maximum} flux
at each wavelength in order  to display  absorption features.  During
Night 2, absorption  at
$-$50 \kms\ 
is replaced by a discontinuous jump to a 
new absorption feature at $-$200 \kms\  (at $\sim$29 hrs).  
{\it Right panel:} Profiles during night 2 of the gray scale representation above.
Note the weakening of the absorption at $-$50 \kms\ during the night, and the 
abrupt appearance of another absorption feature at $-$200 \kms\ in the
middle of the night which  becomes prominent at the end of the night. 
Broad absorption extending from $\sim +$150 to 
$\sim +$400 \kms\ arising from infalling material is  variable, and the velocities
are commensurate with the velocities measured in \ion{He}{1}.}

\end{figure}

\begin{figure}
\includegraphics[angle=90,scale=0.6]{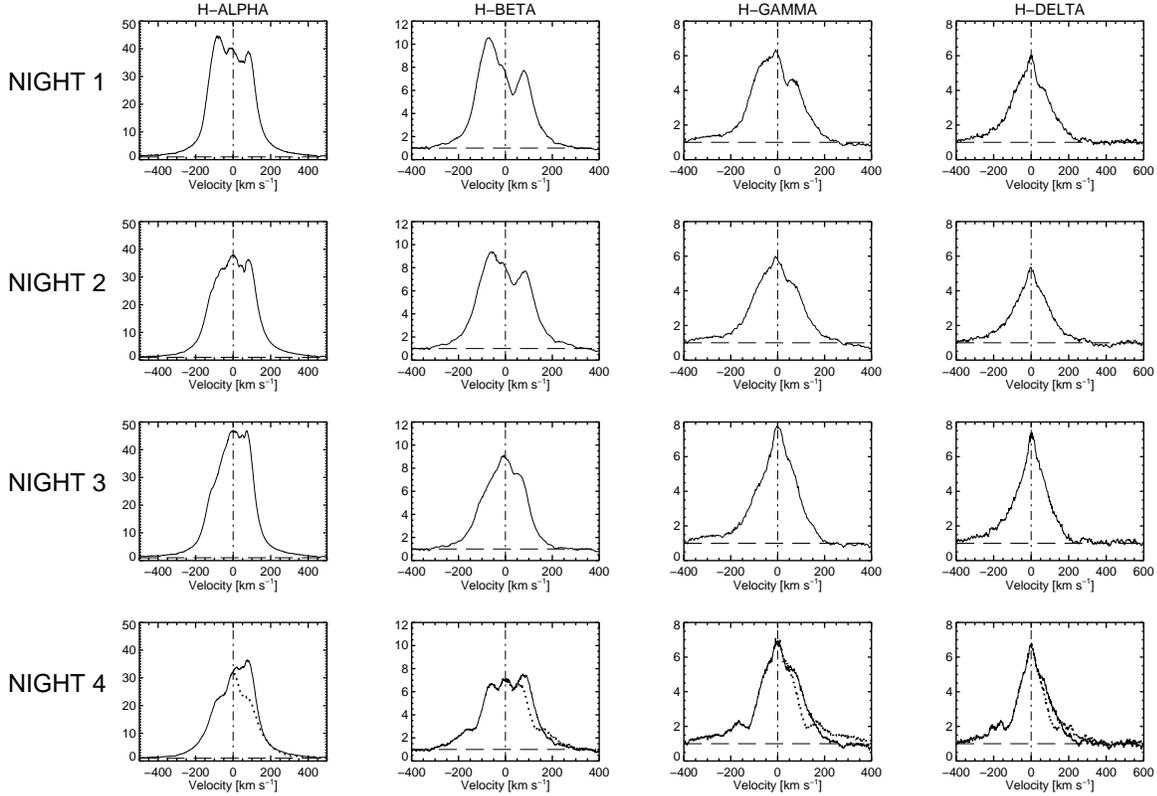}
\caption{Balmer series over 4 consecutive nights in 2007 (Feb. 26 --
  Mar. 1).  The y-axis represents a continuum normalized
  flux. Profiles from night 4 (2007 March 1) have the negative
velocity segments overlaid on the positive velocity side and
marked by a broken ({\it red}) line.  The effects of wind absorption
can be seen when the broken line lies below the solid line;
effects of absorption by the accretion stream can be seen
at $\sim +$200 \kms\ where the broken line lies slightly
above the solid line. }
\end{figure}

\begin{figure}
\includegraphics[angle=90,scale=0.6]{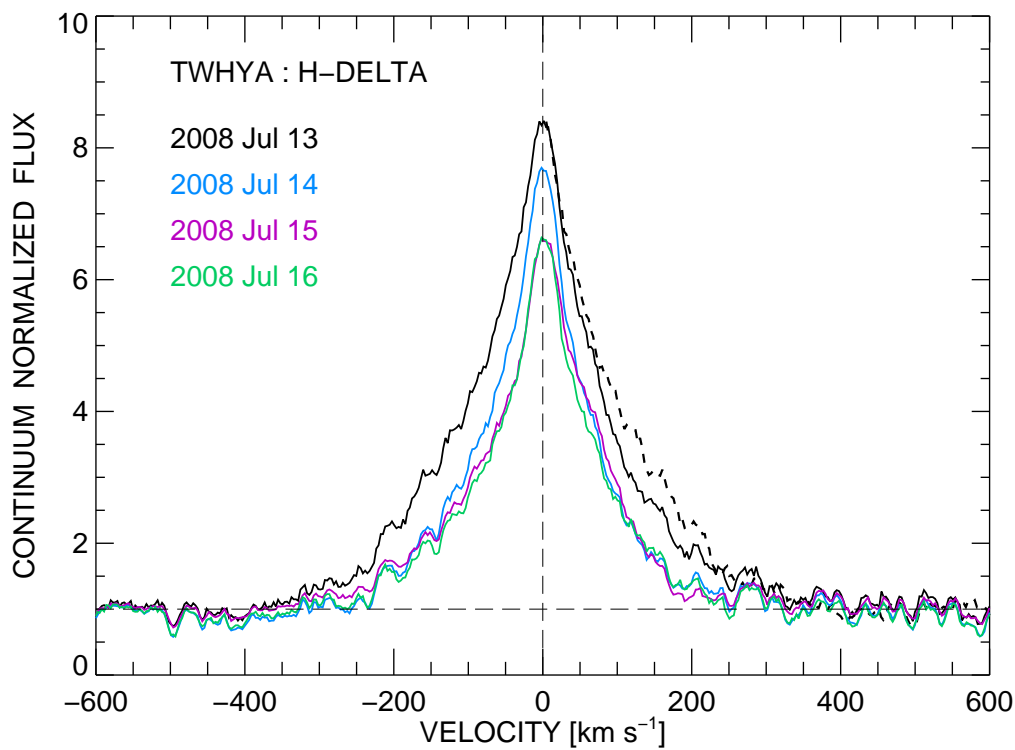}
\caption{Typical H$\delta$ line profile ($\lambda$4101.73) over 4
  consecutive nights in 2008.  The broken line represents the 
negative velocity wing of the first night's observation that has
been reflected around the  axis at zero velocity.  This illustrates the
slight broad absorption present at velocities +50 to +250 \kms.   
Profiles from the following 3 nights are symmetric.  
Subcontinuum absorption predicted by magnetospheric accretion 
models is missing in the positive
velocity wing of the line.}
\end{figure}

\begin{figure}
\begin{center}
\includegraphics[angle=90,scale=0.6]{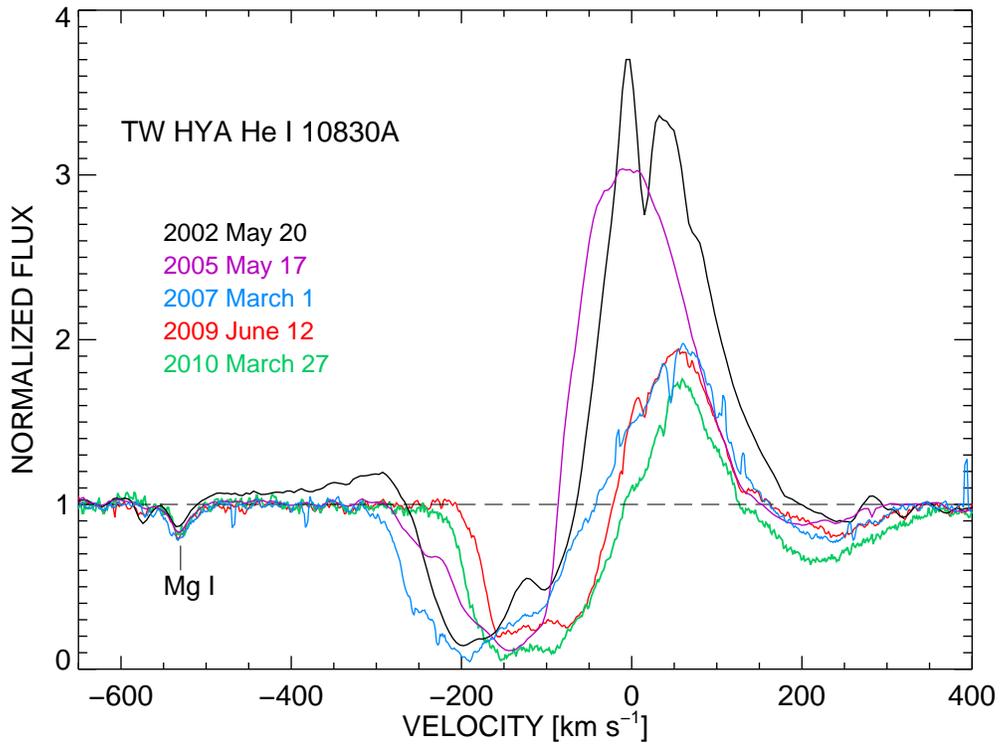}
\caption{Spectra of \ion{He}{1} 10830\AA\ spanning 8 years.
  Substantial changes in the emission flux level, the wind opacity,
  the wind speed, and the inflowing absorption are evident.}
\end{center}
\end{figure}

\clearpage


\begin{figure}
\begin{center}
\includegraphics[angle=0., scale=0.45]{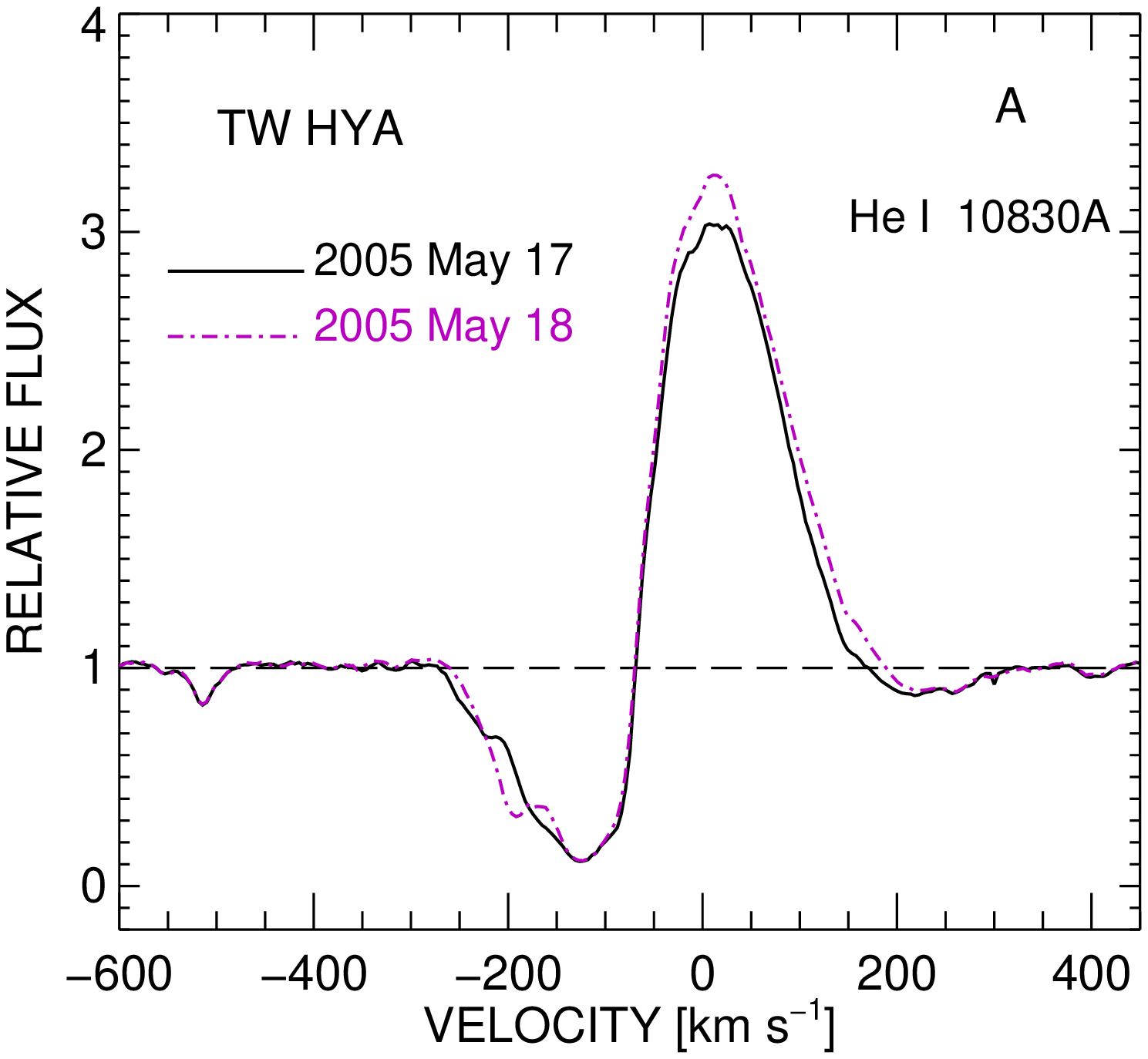}
\includegraphics[angle=0,scale=.45]{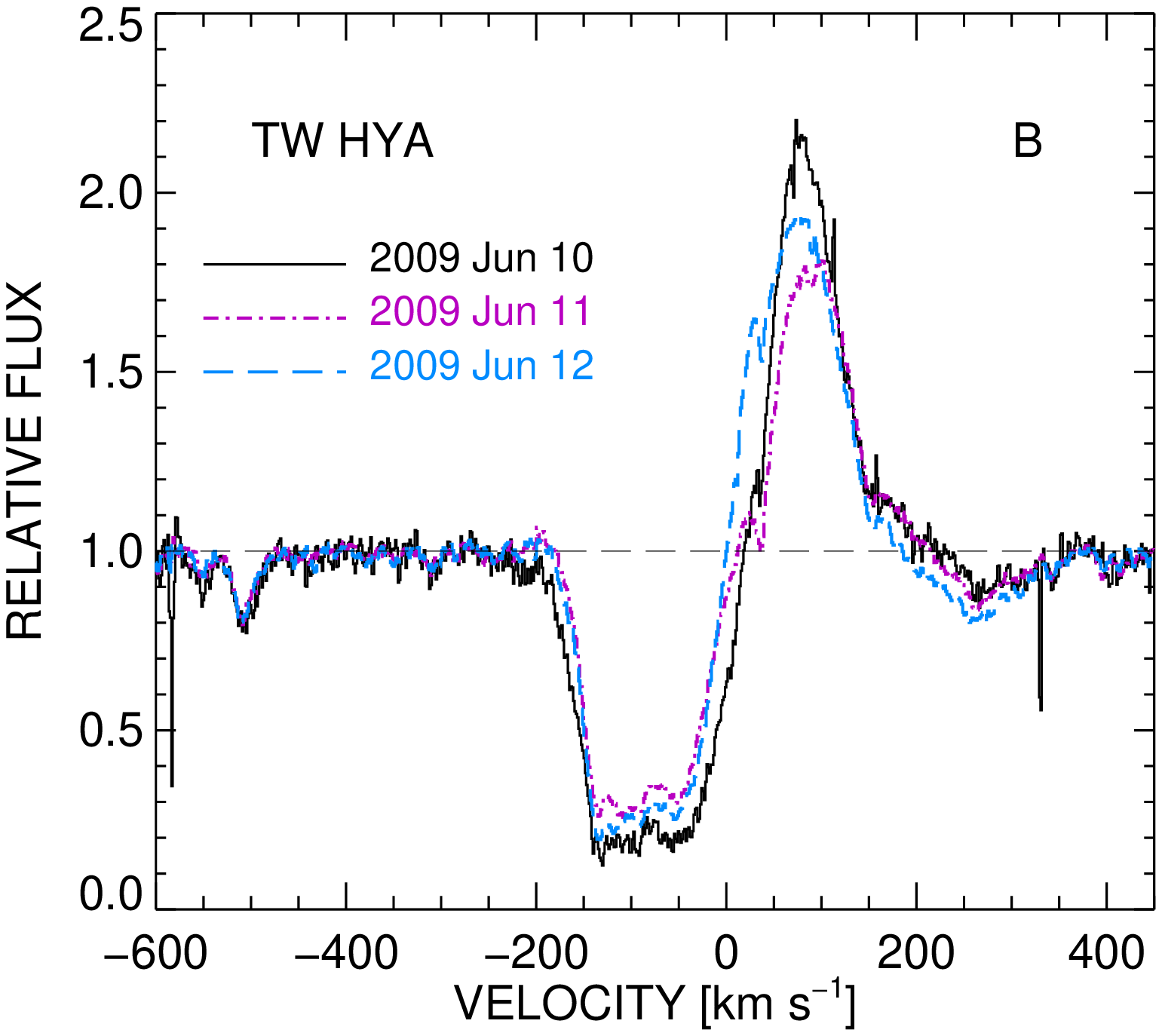}
\includegraphics[angle=0,scale=.45]{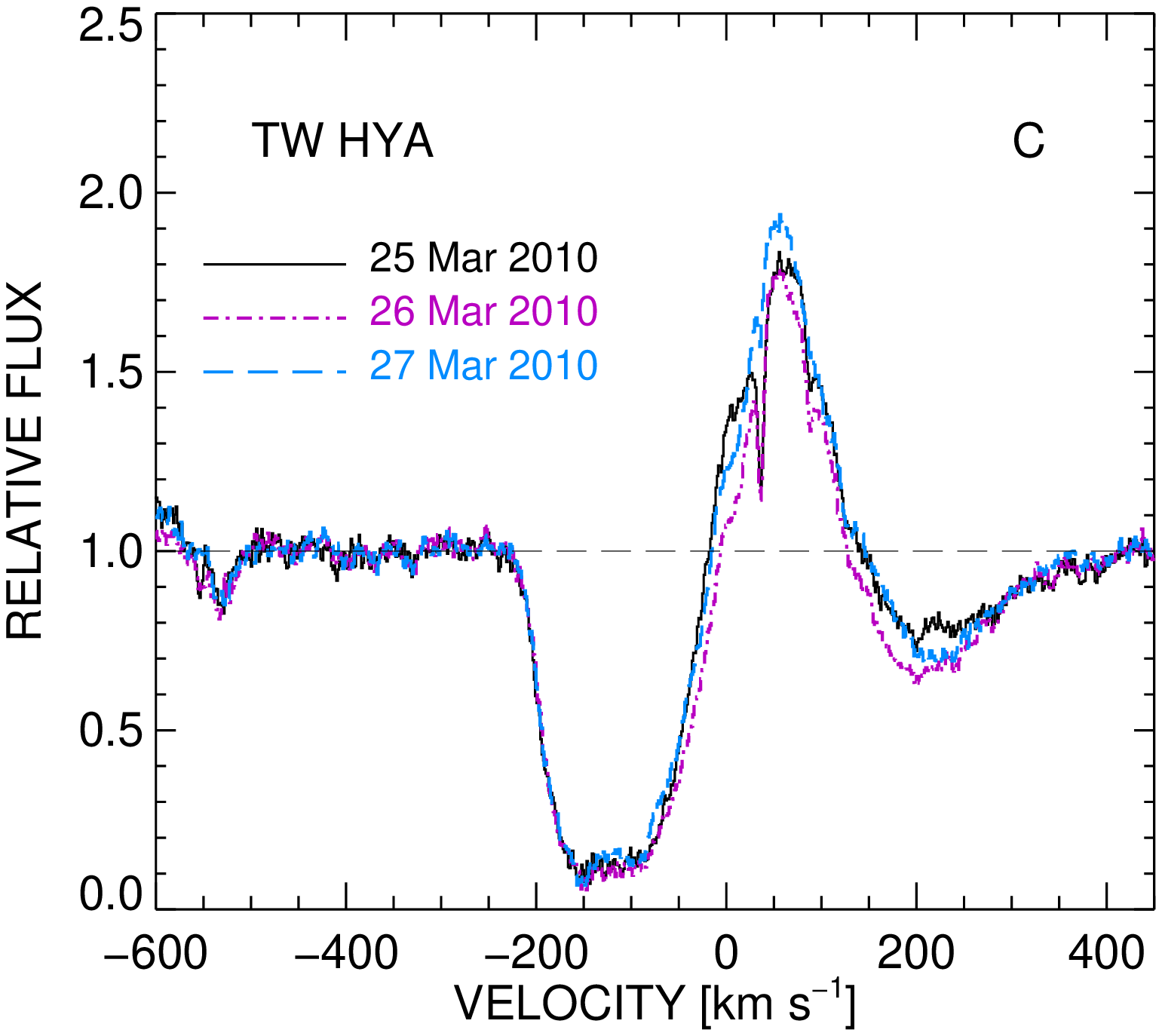}
\caption{KECK II NIRSPEC spectra (2005) and Gemini-S PHOENIX spectra
  (2009, 2010) of TW Hya on  successive nights.  The outflowing wind
remains relatively stable as compared to both the emission
and the inflowing material.}
\end{center}
\end{figure}


\begin{figure}

\hspace*{-0.8in}

\includegraphics[angle=0,scale=0.5]{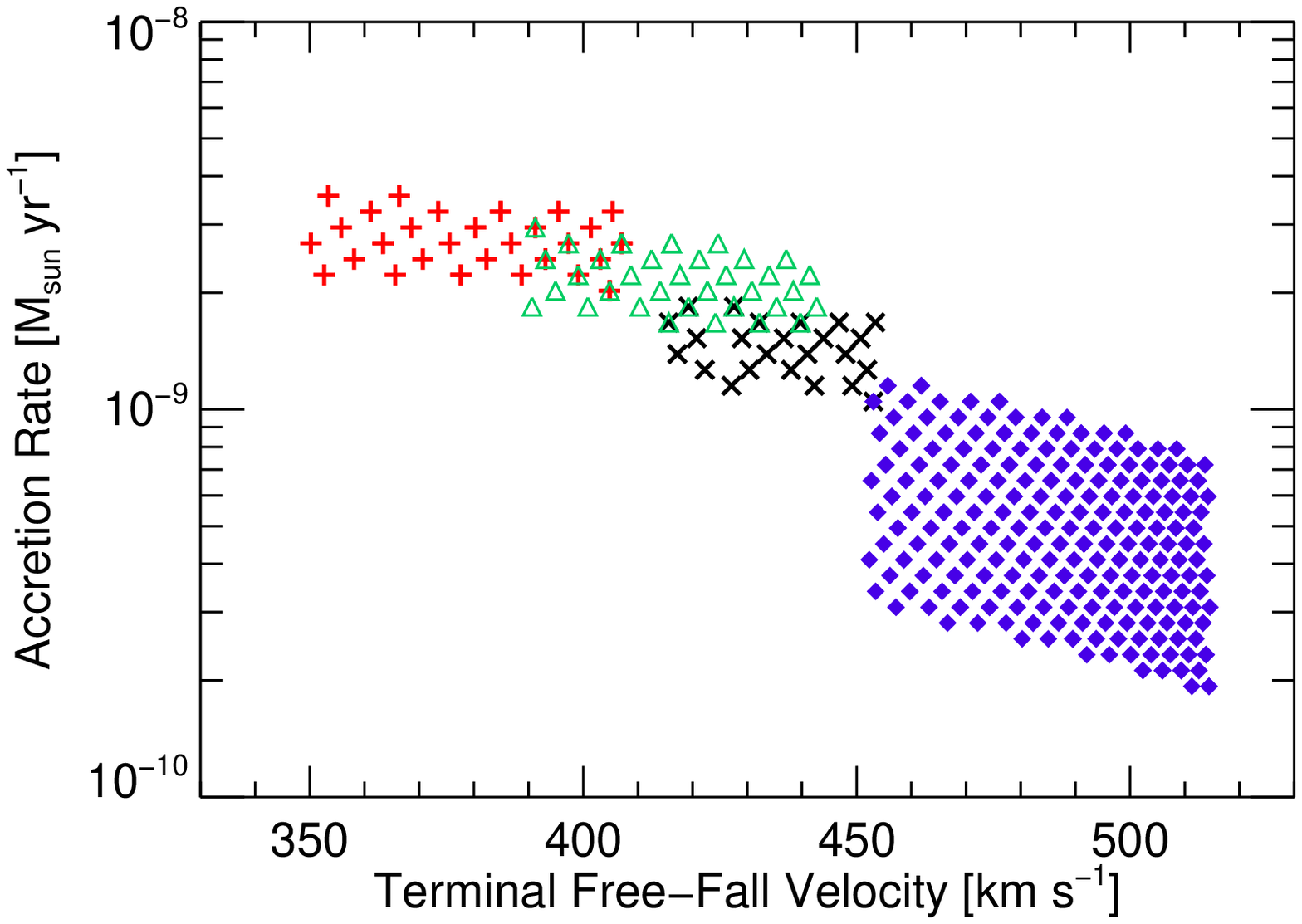}
\vspace*{-2.5in}

\hspace*{3.3in}
\includegraphics[angle=0.,scale=0.5]{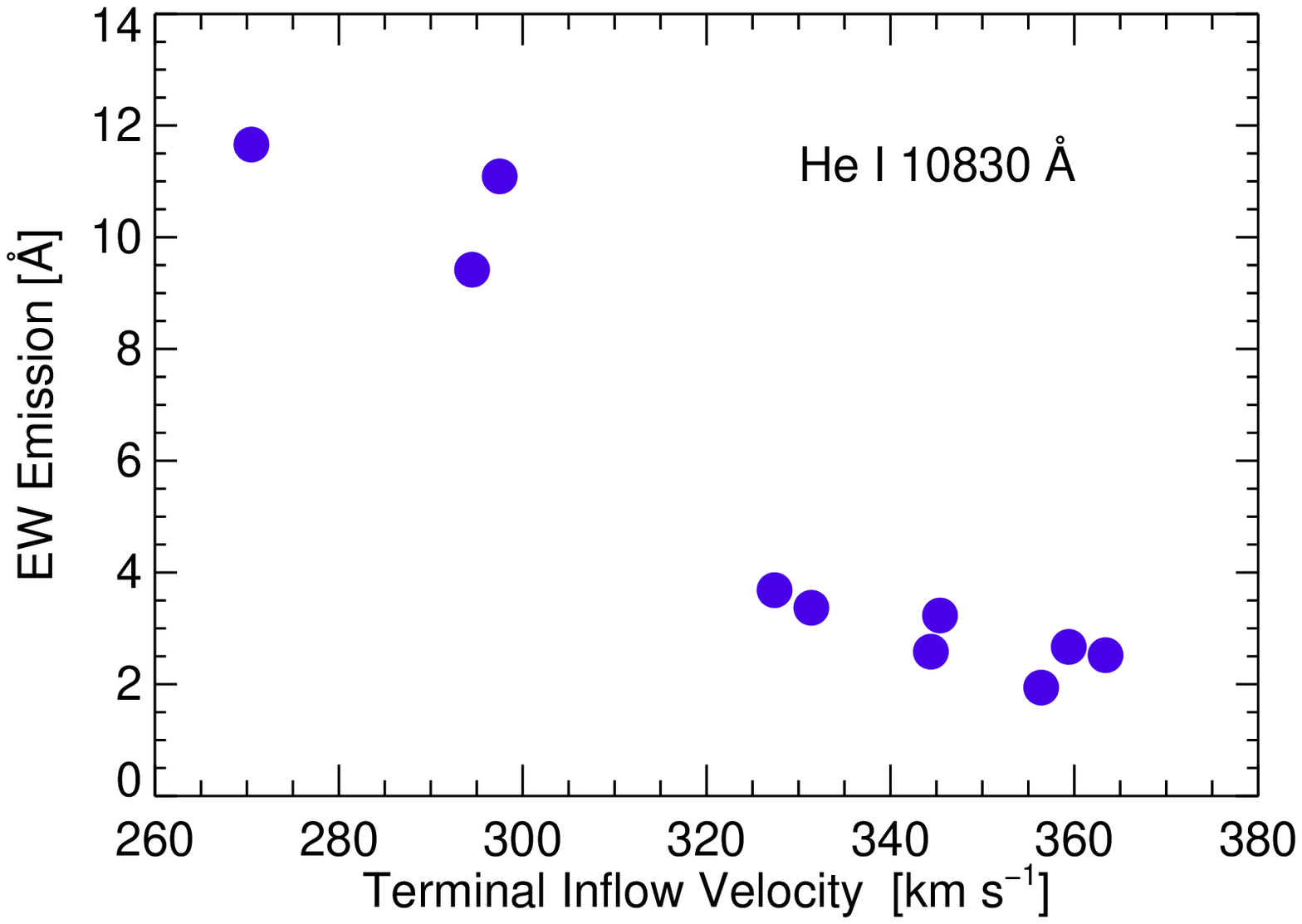}

\caption{{\it Left panel:}     The magnetic dipole accretion
model  predicts the relationship between the accretion rate
and the terminal free-fall velocity.  A 
slower infall  velocity results from a smaller radius of the inner
circumstellar disk, and a larger filling factor in the stellar atmosphere 
produces  a higher mass accretion rate. Parameter ranges
constrained
by CHANDRA \ion{Ne}{9} diagnostics are shown here.  The  $\times$ symbols
denote the average CHANDRA spectrum, while the other symbols mark 3 individual
CHANDRA pointings (cf. Brickhouse \etal\ 2012).  {\it Right panel:}
Equivalent width of the emission component of the \ion{He}{1} line
($\lambda$10830) as a function of the terminal velocity indicated by the
inflowing subcontinuum absorption.  These observations display general
agreement with the magnetic dipole accretion model. The emission in
the post-shock cooling zone increases because of increased accretion
at lower terminal free-fall velocities.}

\end{figure}


\begin{figure}
\begin{center}
\includegraphics[angle=0,scale=0.7]{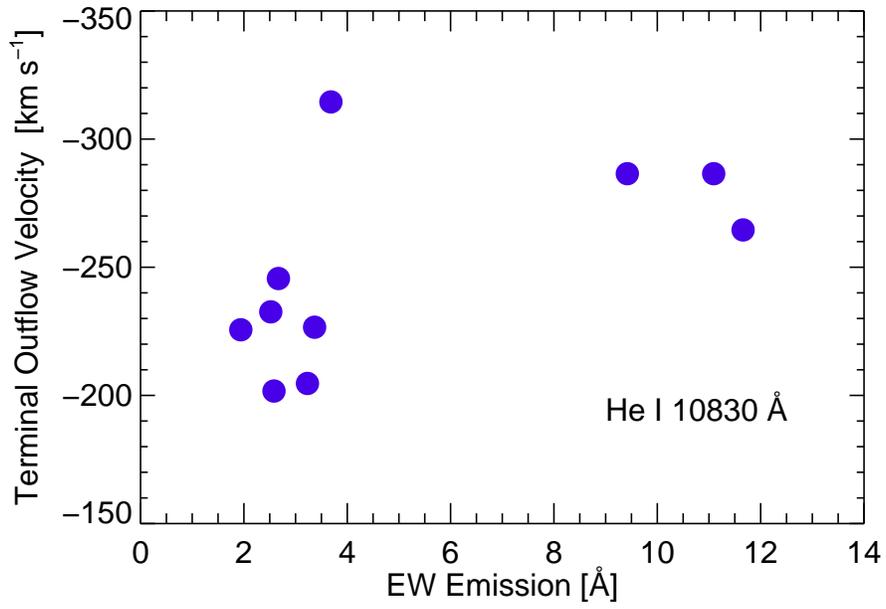}
\caption{The terminal outflow velocity as a function of the
emission component of the \ion{He}{1}. Except for one outlier (2007 Mar. 1),
fast outflow is correlated with increased emission. This is not
inconsistent with  the conclusion that the 
turbulent post-shock region can affect the wind speed. }
\end{center}
\end{figure}

\begin{figure}
\includegraphics[angle=0.,scale=0.5]{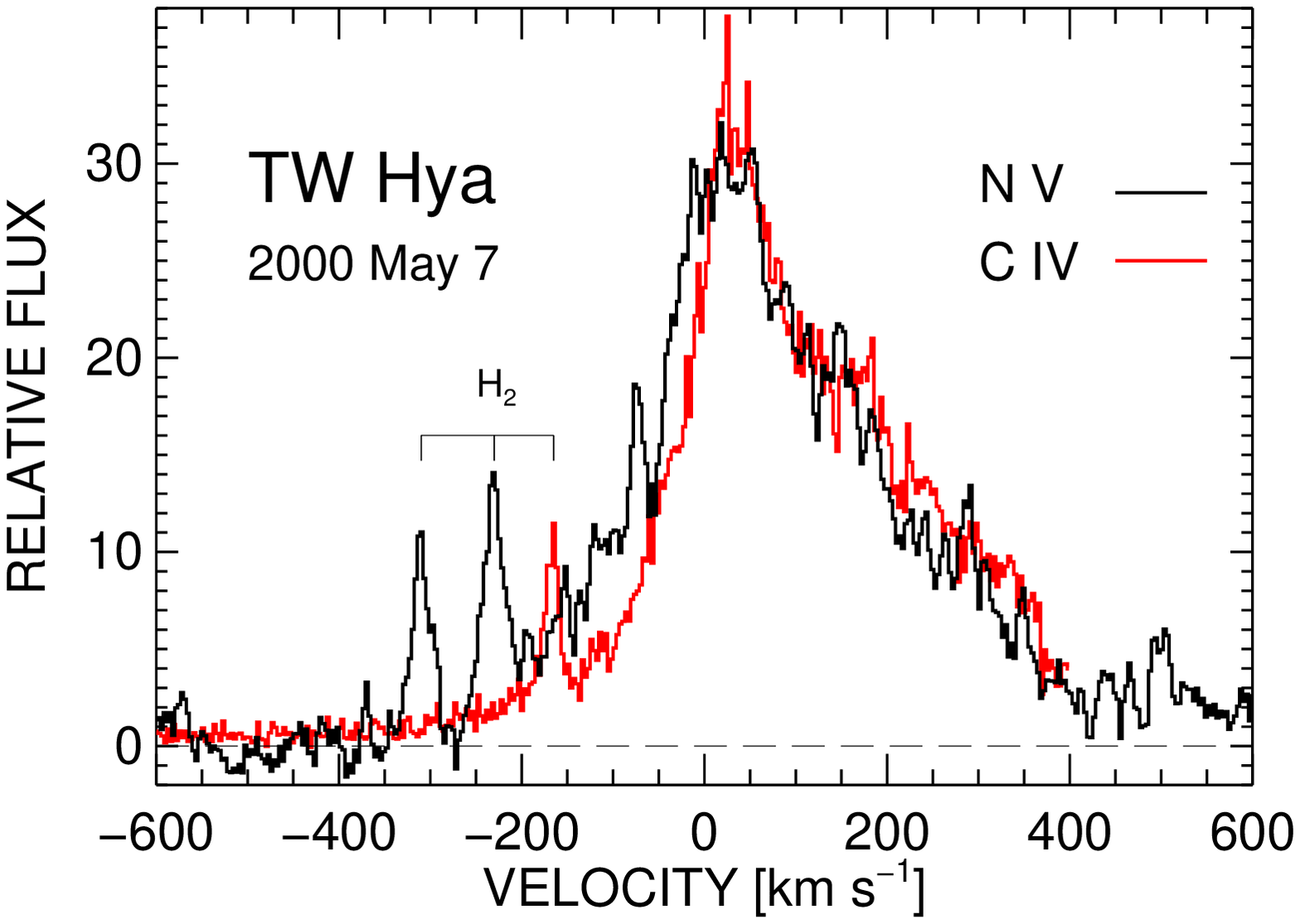}
\vspace*{-2.5in}

\hspace*{3.5in}
\includegraphics[angle=0.,scale=0.5]{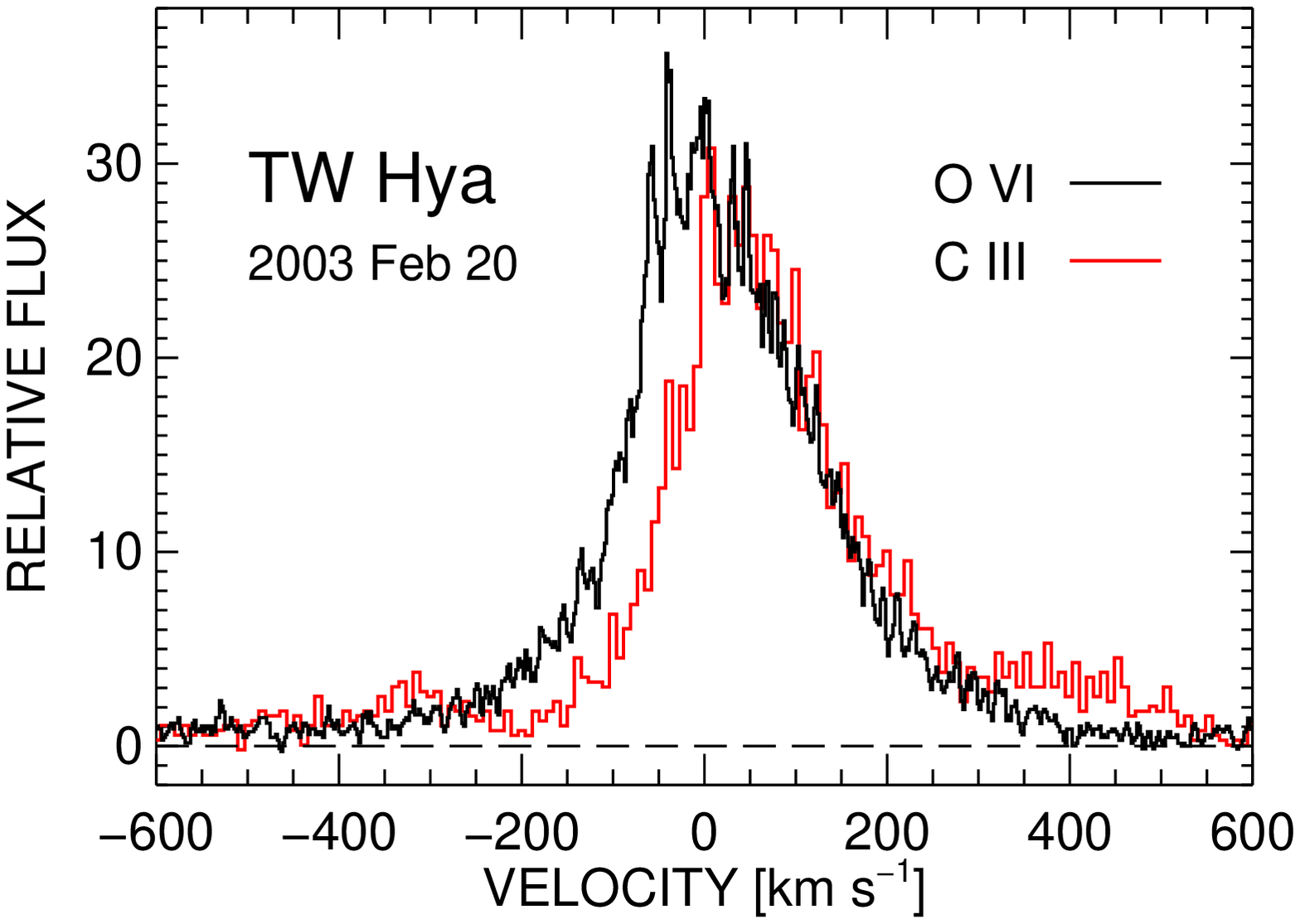}
\caption{UV and far-UV spectra of major resonance lines. Peak fluxes are
scaled to match to  illustrate the shape of the profiles. {\it Left panel:} \hst:STIS
spectra of \ion{N}{5} (\gla 1239) and \ion{C}{4}(\gla 1548) taken in the year
2000. {\it Right panel:} \fuse\  spectra of \ion{C}{3} (\gla 977) and
\ion{O}{6} taken three years later in 2003 (Dupree \etal\ 2005a).  Note that the profiles
observed at the same time are similar one to another with the
exception of
different absorption levels on the negative velocity side (see text for explanation).} 
\end{figure}


\begin{figure}
\begin{center}
\includegraphics[angle=0., scale=0.6]{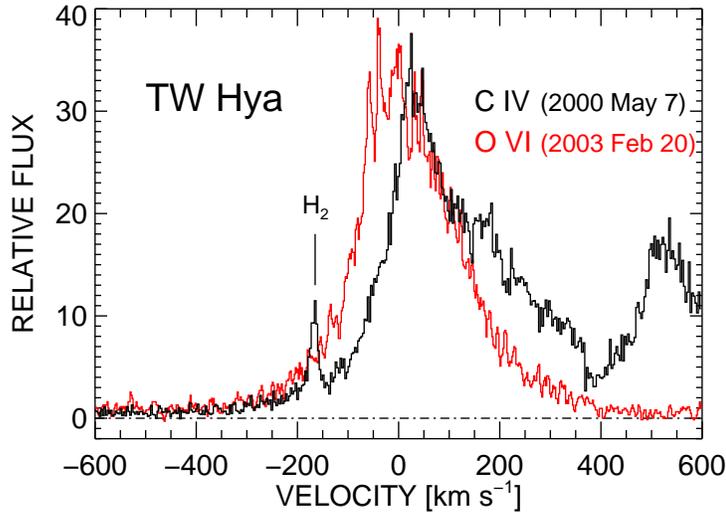}
\end{center}
\caption{Comparison of resonance lines at different times showing
the dramatic change in emission at positive velocities.  The second
member of the \ion{C}{4} doublet appears at $+$500 \kms. See text
for discussion of these profiles.   
Peak fluxes are scaled in this figure to illustrate the shape of the line profiles.}
\end{figure}

\begin{figure}
\includegraphics[angle=0.,scale=0.5]{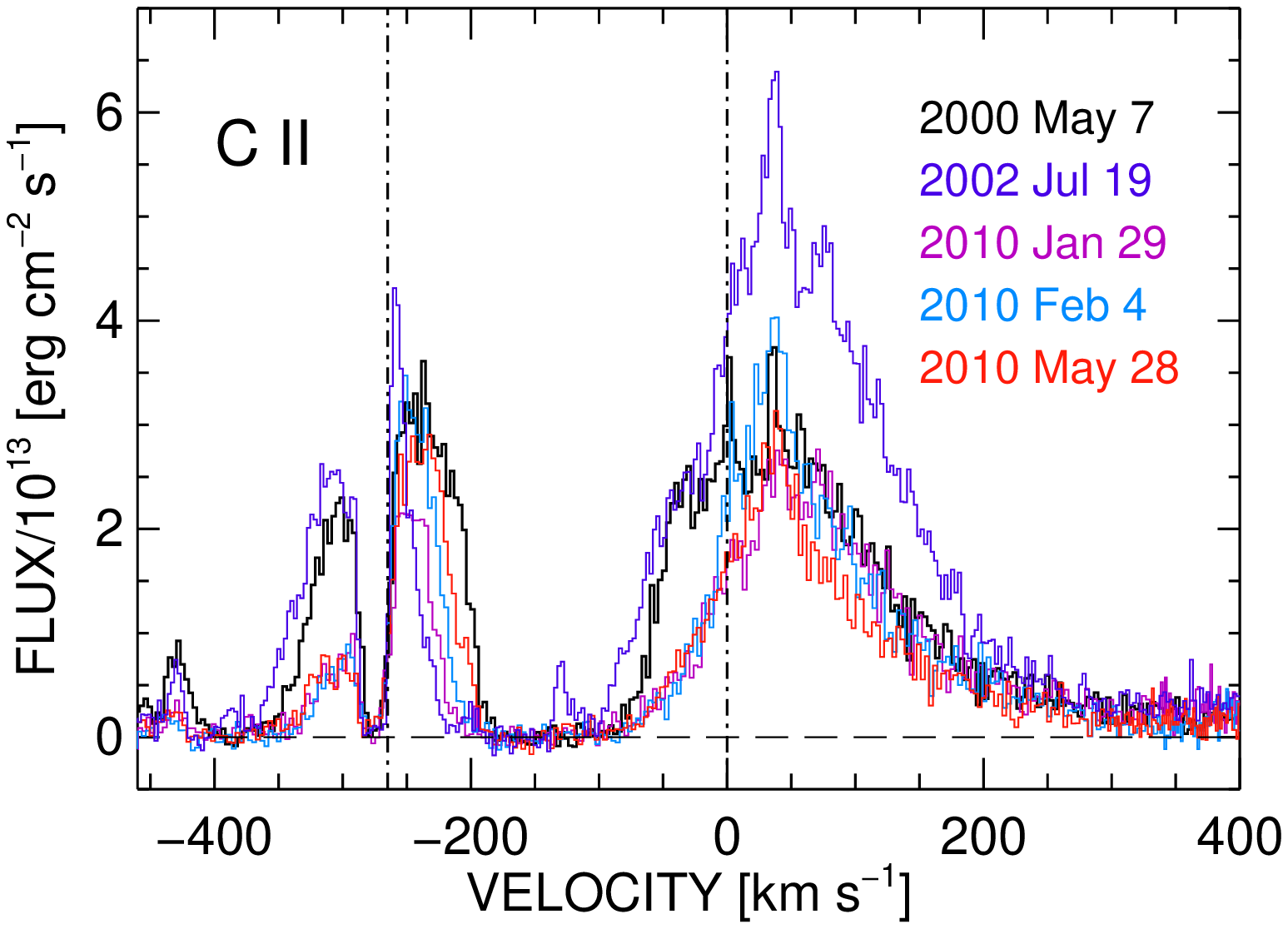}
\vspace*{-2.5in}

\hspace*{3.2in}
\includegraphics[angle=0.,scale=0.5]{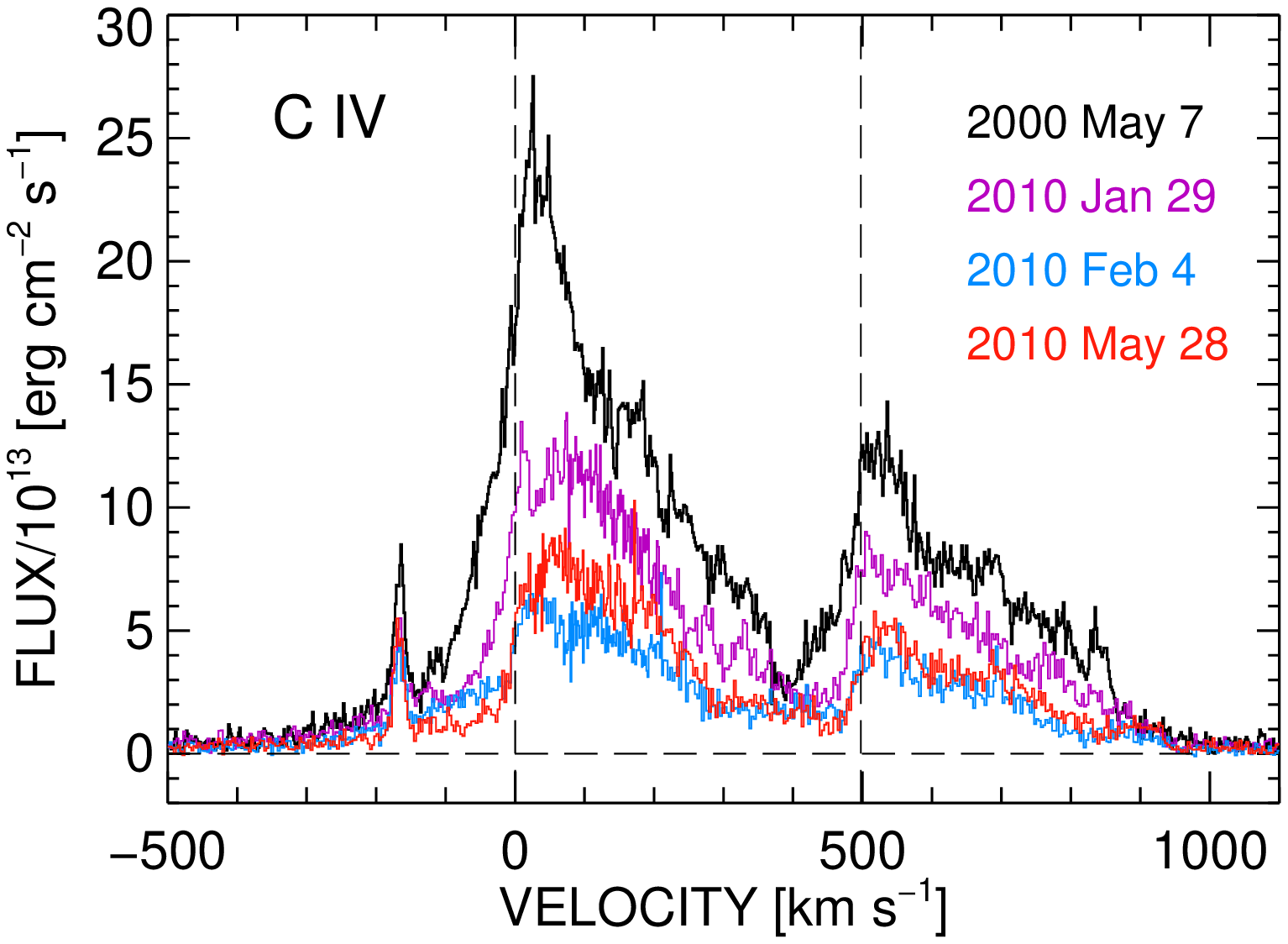}
\caption{\ion{C}{2} ({\it 1335\AA}) and \ion{C}{4} ({\it 1548\AA})
  multiplets showing 
the changes in both intrinsic line strength and
    wind opacity, even on short time scales, such as the 6 days
    separating 2010  Jan 29  and 2010 Feb 4. The \ion{C}{2} line is generally
   narrower on the positive velocity side than the \ion{C}{4} which
is similar to the high temperature \ion{N}{5} profile. The broken
    vertical lines mark the rest velocities of the doublets.}
\end{figure}


\begin{figure}
\includegraphics[angle=0.,scale=0.5]{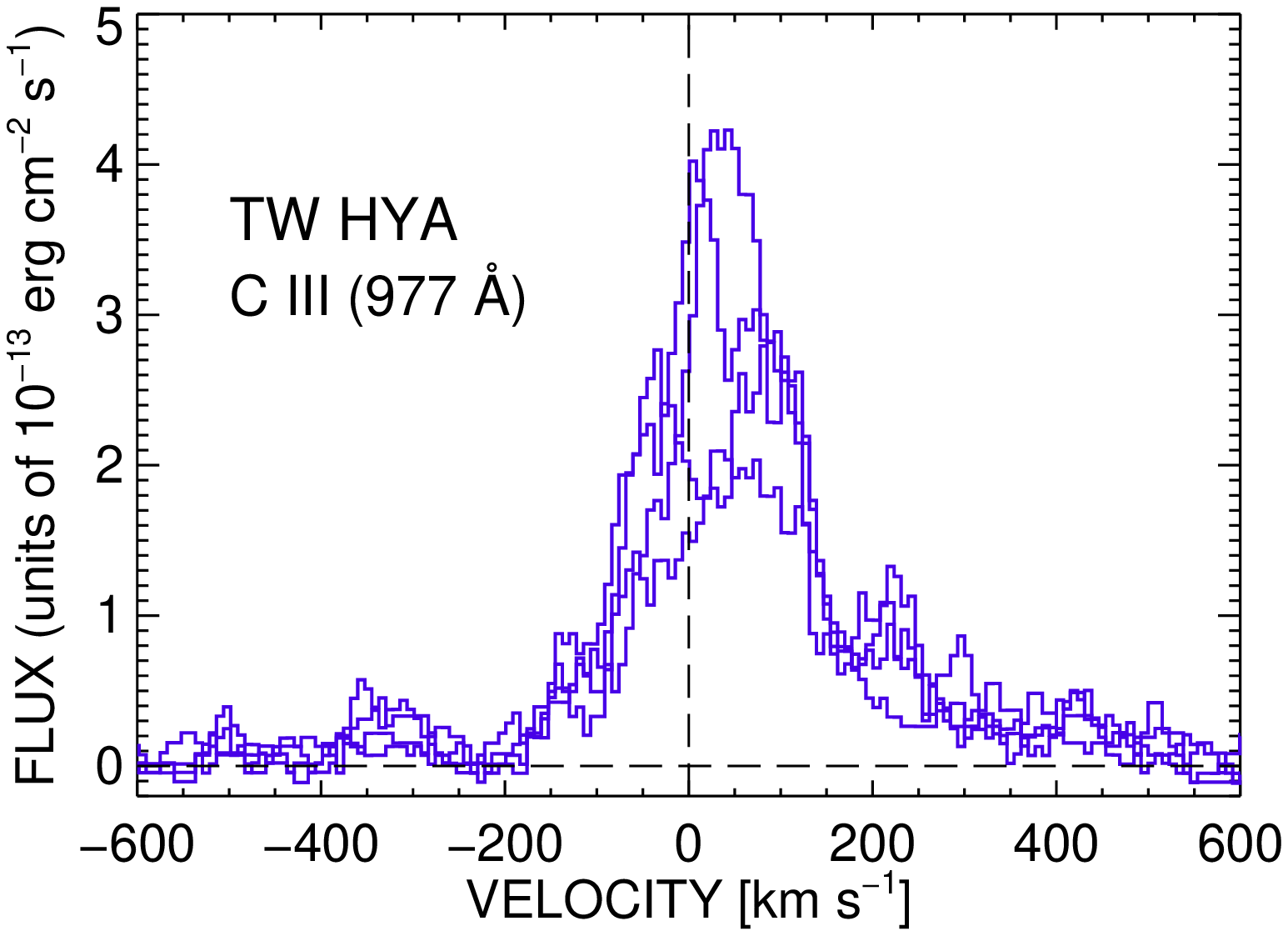}

\vspace*{-2.53in}

\hspace*{3.3in}
\includegraphics[angle=0.,scale=0.5]{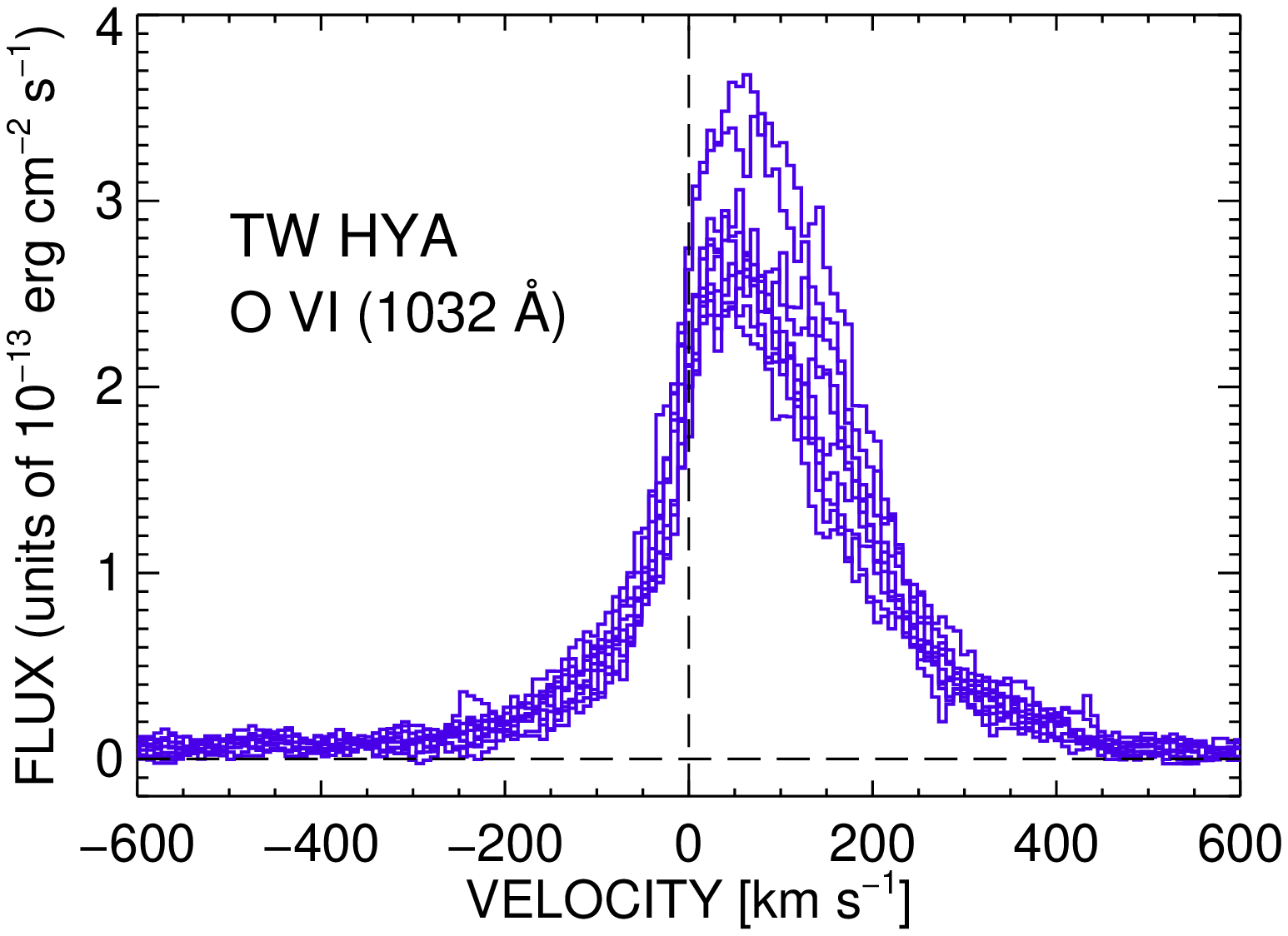}

\caption{FUSE time-tagged spectra of \ion{C}{3} (977\AA) and  O VI
  (1031.91\AA) taken through the LWRS aperture from the
 SiC2A and LiF1A detector measured over a 32.2 hour span 2003 Feb 20-21.   The
 variability at positive velocities is more substantial in both lines than at
negative velocities.  The opacity of the hot wind as measured in the \ion{O}{6} line remains
  relatively constant.  Flux measurements from the individual spectra are shown in Figure 16.
  }

\end{figure}

\begin{figure}
\begin{center}

\includegraphics[angle=0.,scale=0.6]{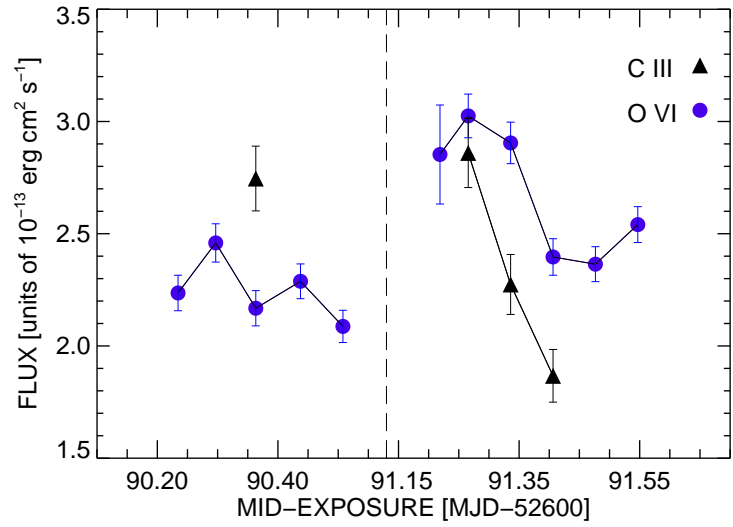}

\caption{Fluxes measured from the {\it FUSE}\  line profiles shown in the previous
  figure display similar behavior in the \ion{C}{3} and \ion{O}{6} emission during several
hours of the second pointing. During the second pointing, the absolute flux values
are reversed from the first pointing, and the line fluxes are correlated.} 

\end{center}
\end{figure}


\begin{figure}
\begin{center}

\vspace*{-0.3in }
\includegraphics[angle=90.,scale=0.5]{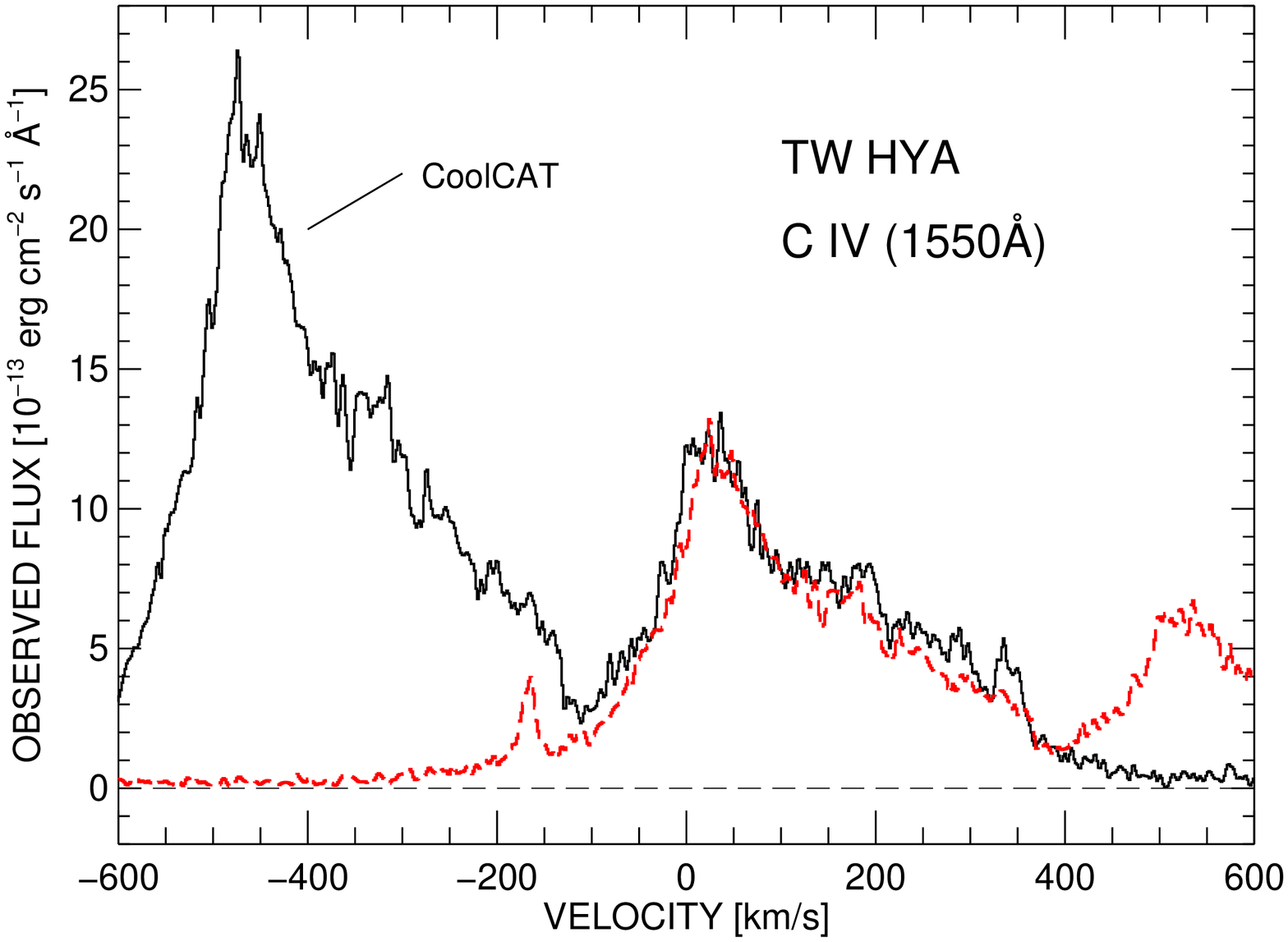}

\vspace*{-0.1in}

\includegraphics[angle=90.,scale=0.5]{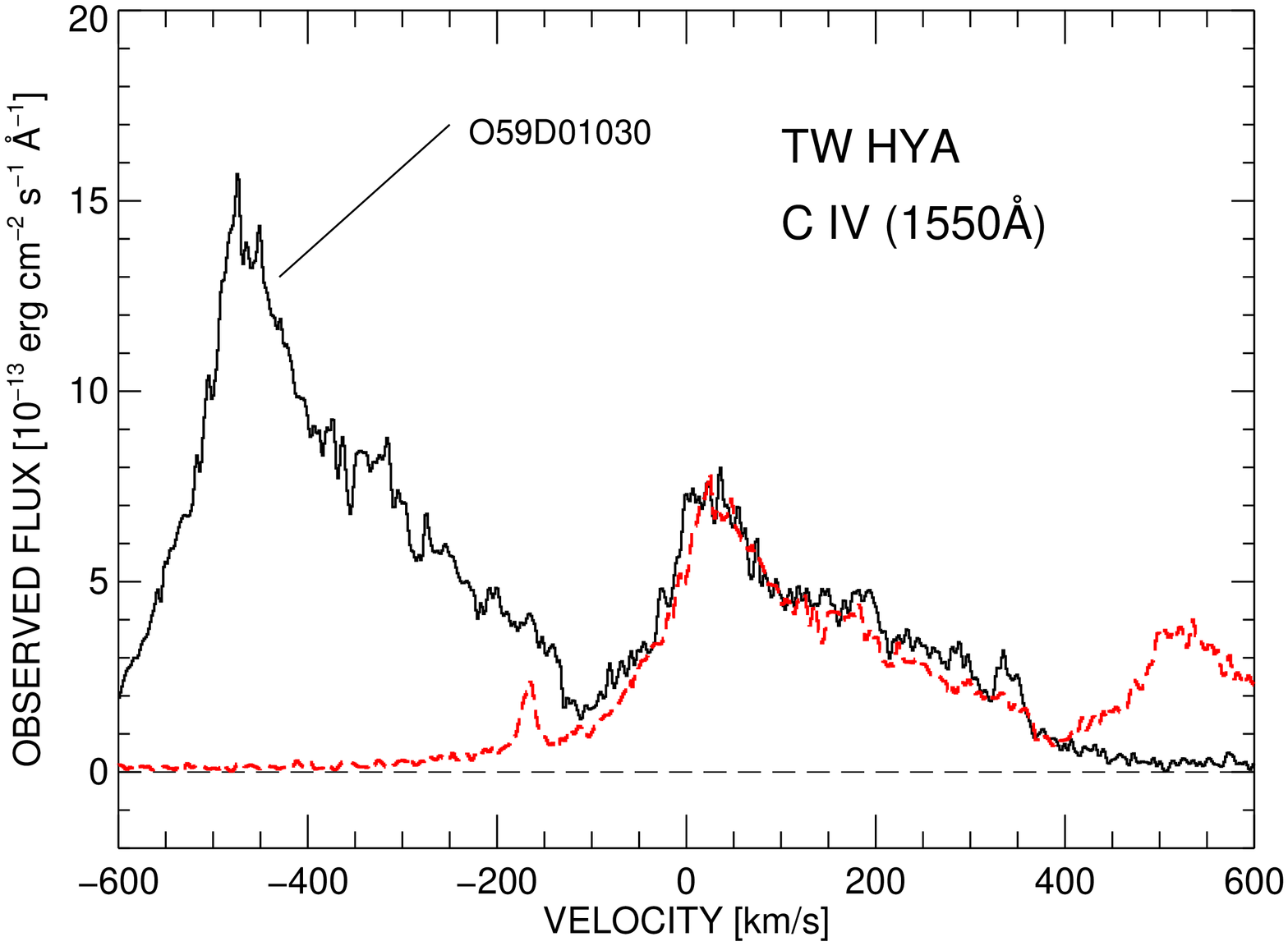}
\caption{The solid  lines show the `CoolCAT' optimum reduction ({\it
    upper panel}) of the same STIS  spectrum as that used by JKH 
({\it lower panel}, O59D01030).    
The broken  lines (red in online version) represent the profile
of the 1548\AA\ multiplet multiplied by 0.5 and shifted by
the velocity equivalent of the wavelength separation of the
doublet. Wind absorption by the 1550\AA\ component is 
 indicated by the weaker extended positive velocity 
wing of the 1548\AA\ transition ({\it broken line}). 
Wind absorption occurs systematically between $+$175 and  
$+$375 \kms\ of the scaled 1548\AA\ line in the 
velocity scale of the figure.    Additionally the effect of
 increased opacity in the 1548\AA\ line  
is evident in the region $-$125 to $+$25 \kms, where 
the broken line lies systematically below the solid line.  Thus, two signatures of
wind absorption can be noted in these doublet profiles: absorption
of the red wing of 1548\AA\  by the blue wing of 1550\AA, and the 
increased scattering of the 1548\AA\  line in the wind as compared with 1550\AA.}
\end{center}
\end{figure}

\begin{figure}
\begin{center}
\includegraphics[scale=0.8]{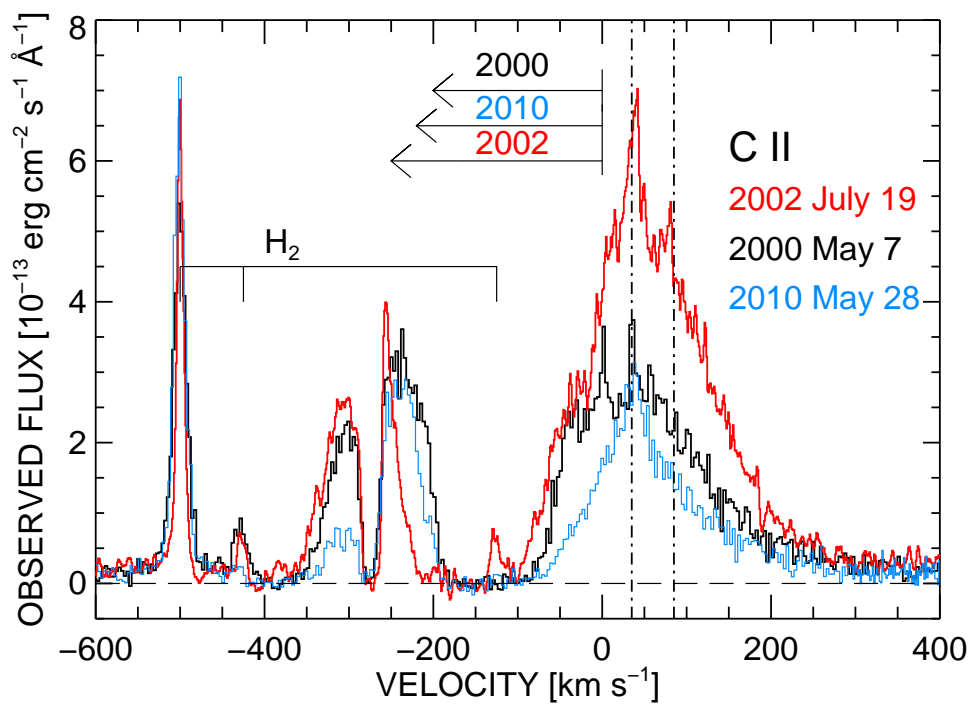}

\caption{HST/STIS spectra of TW Hya in the region of the \ion{C}{2}
 1335\AA\  doublet in 2000, 2002, and 2010 showing the variabilities
in \ion{C}{2} emission, H$_2$ emission and wind speed and opacity.  
The short wavelength component
 ($-$275 km s$^{-1}$) of the doublet is
  bifurcated by interstellar absorption.
The broken vertical lines  mark the region ($+$35 to $+$85 \kms)  where JKH postulated
the lack of absorption.  This section of the profile is obviously
substantially variable in both flux and slope of the emission line.
 The arrows indicate the extent of wind absorption for one member of
 the doublet.  Note that the $H_2$ lines vary, but not in any
systematic way with wind speed or opacity. 
In 2002, when the wind opacity is substantial  between $-$100 and
 $-$250 \kms,  another $H_2$ emission line
 appears (R(2)0-4 at 1335.2\AA\  at $\sim -$125 \kms).
 }
\end{center}
\end{figure}

\begin{deluxetable}{lllll}
\def\a{\phantom{0}}
\def\b{\phantom{00}}
\tablecolumns{5}
\tablewidth{0pt}
\tablenum{1}
\tablecaption{TW Hya Observations}
\tablehead{
\colhead{Spectral}   &     
\colhead{Instrument}  &     
\colhead{Date}  &
\colhead{Data file/} &
\colhead{Exp. time } \\
\colhead{Region}&
\colhead{} &
\colhead{} &
\colhead{spectra} &
\colhead{(s)} 
 }
\startdata
Far UV& FUSE & 2003  Feb 20 & C0670101\tablenotemark{a}& 15221    \\
   & FUSE    & 2003 Feb 21 & C0670102\tablenotemark{a}& 15465 \\
UV & HST/STIS/E140M  &2000 May 7 &O59D01030\tablenotemark{b}& 2300   \\
 &HST/STIS/E140H&  2002 July 19 & StarCAT\tablenotemark{c} & 4355 \\
 & HST/STIS/E140M &2010 Jan 29 & OB3R07060\tablenotemark{d}&3058\\
 & HST/STIS/E140M &2010  Feb  04 & OB3R08060\tablenotemark{d}&3058\\
& HST/STIS/E140M  &2010 May 28 & OB3R09060\tablenotemark{d}&3058\\
Optical &Magellan/CLAY/MIKE& 2004 Apr 27--30&19& 60-300     \\
 &Magellan/CLAY/MIKE& 2005 Jul 24--27&4& 60-300     \\
 &Magellan/CLAY/MIKE& 2006 Apr 15--17&19& 60-300     \\
 &Magellan/CLAY/MIKE& 2007 Feb 26--Mar 1&$\sim$360& 45-360     \\
&Magellan/CLAY/MIKE& 2007 Jul 23--25&4&45-60   \\  
&Magellan/CLAY/MIKE& 2008 May 25--26&7&45-240   \\
 &Magellan/CLAY/MIKE& 2008 Jul 13--16&9&35-180 \\
 &Magellan/CLAY/MIKE& 2009 May 4--5&5&45-180 \\
  &FLWO/TRES& 2009 Jun 10--12&3&720    \\
  &Magellan/CLAY/MIKE& 2010 Jun 30--Jul 2&3& 45\\
  &FLWO/TRES& 2011 Apr 11--15& 5&1200 \\
 &Magellan/CLAY/MIKE& 2013 Jul 19--25 & 5&30-180\\
Near IR &KECK II/NIRSPEC& 2002 May 20&2&200\\
 &KECK II/NIRSPEC& 2005 May 17--18&4&200\\
 &Gemini-S/PHOENIX& 2007 Mar 1 &1&900 \\ 
   &Gemini-S/PHOENIX& 2009 Jun 10-12 & 3 & 300 \\
        &Gemini-S/PHOENIX& 2010 Mar 25--27 & 3 & 300\\
\enddata
\tablenotetext{a}{PI: J. Linsky}
\tablenotetext{b}{CoolCAT
({\it casa.colorado.edu/\ $\tilde{ }$ayres/CoolCAT}) was derived from this spectrum.}
\tablenotetext{c}{StarCAT can be found at {\it
    archive.stsci.edu/prepds/starcat} (Ayres 2010).}
\tablenotetext{d}{PI: N. Calvet}
\end{deluxetable}


\end{document}